\documentclass[usenatbib]{mnras}

\usepackage{natbib}
\usepackage{delarray}
\usepackage{color}
\usepackage{amsmath}
\usepackage{amssymb}
\usepackage{here}
\usepackage{graphicx}
\usepackage[utf8]{inputenc}
\usepackage{float}
\usepackage{epsfig}
\usepackage{tensor}
\usepackage{xspace}

\usepackage{newtxtext,newtxmath}

\usepackage[T1]{fontenc}
\usepackage{ae,aecompl}

\newcommand{\appropto}{\mathrel{\vcenter{
  \offinterlineskip\halign{\hfil$##$\cr
    \propto\cr\noalign{\kern2pt}\sim\cr\noalign{\kern-2pt}}}}}

\newcommand{\mach}{\mathcal{M}}

\newcommand{\be}{\begin{equation}} \newcommand{\ee}{\end{equation}}

\newcommand{\lbra}{\left(}
\newcommand{\rbra}{\right)}
\newcommand{\dderiv}{\mathrm{d}}
\newcommand{\vvector}{\mathbf{v}}

\newcommand{\acknowledgments}{\begin{small}\section*{Acknowledgments}\end{small}}
\newcommand\altaffilmark[1]{$^{#1}$}
\newcommand\altaffiltext[1]{$^{#1}$}

\newcommand{\myquote}[1]{``#1''}

\usepackage[normalem]{ulem}

\title[Isothermal Fragmentation]{Isothermal Fragmentation: Is there a low-mass cut-off?}

\author[Guszejnov, Hopkins, Grudi{\'c}, Krumholz  \&\  Federrath]{
\parbox[t]{\textwidth}{ D\'avid Guszejnov\altaffilmark{1}\thanks{E-mail:guszejnov@caltech.edu}, Philip F. Hopkins\altaffilmark{1}, Michael Y. Grudi{\'c} \altaffilmark{1}, Mark R. Krumholz\altaffilmark{2,3} and Christoph Federrath\altaffilmark{2}}
\vspace*{6pt} \\
\altaffiltext{1}{TAPIR, MC 350-17, California Institute of Technology, Pasadena, CA 91125, USA} \\
\altaffiltext{2}{Research School of Astronomy and Astrophysics, The Australian National University, Canberra, ACT 2611 Australia}\\
\altaffiltext{3}{ARC Centre of Excellence for All Sky Astrophysics in 3 Dimensions (ASTRO 3D), Australia}
}

\date{To be submitted to MNRAS, \today \vspace{-0.6cm}}
\begin{document}
\maketitle
\label{firstpage}

\begin{abstract}
The evolution of self-gravitating clouds of isothermal gas forms the basis of many star formation theories. Therefore it is important to know under what conditions such a cloud will undergo monolithic collapse into a single, massive object, or will fragment into a spectrum of smaller ones. And if it fragments, do initial conditions (e.g. Jeans mass, sonic mass) influence the mass function of the fragments, as predicted by many theories of star formation? In this paper we show that the relevant parameter separating monolithic collapse from fragmentation is not the Mach number of the initial turbulence (as suspected by many), but the infall Mach number $\mach_{\rm infall}\sim\sqrt{G M/(R c_s^2)}$, equivalent to the number of Jeans masses in the \emph{initial} cloud $N_J$. We also show that fragmenting clouds produce a power-law mass function with slopes close to the expected -2 (i.e. equal mass in all logarithmic mass intervals). However, the low-mass cut-off of this mass function is entirely numerical; the initial properties of the cloud have no effect on it. In other words, if $\mach_{\rm infall}\gg 1$, fragmentation proceeds \emph{without limit} to masses much smaller than the initial Jeans mass.
\end{abstract}

\begin{keywords}
stars: formation -- turbulence  -- cosmology: theory -- hydrodynamics
\vspace{-1.0cm}
\end{keywords}
\section{Introduction}\label{sec:intro}

The evolution of a gravitationally bound isothermal fluid is the \myquote{base model} for a large number of astrophysical phenomena, including the formation of stars. In the case of star formation the highly efficient cooling of molecular gas produces an approximately isothermal behaviour on a wide range of scales. Of course this neglects a huge range of physics (e.g. radiation, magnetic fields, optically thick cooling), but clearly one would like to understand this nominally simple case before considering additional physics.

The first modern theories of star formation showed that dense gas clouds are unstable to gravitational collapse \citep{Jeans_1902}, which still forms the basis of our understanding of the process. Later analytical work showed that (in highly idealized scenarios) the characteristic length scale of the instability (Jeans-length) decreases faster than the original cloud, leading to fragmentation \citep[see e.g.][]{Hoyle_1953,Hunter_1962,Hunter_1964}, which would repeat in the (idealized) evolution of these substructures. Later \cite{Tohline_1980} showed that it is actually possible for an isothermal cloud to collapse to a single object without fragmenting. This is a key assumption of the modern \myquote{gravito-turbulent} star formation models \citep[e.g.][]{Padoan_Nordlund_2002_IMF,HC08,HC2009,HC_2013,excursion_set_ism}. These models predict the mass distribution of self-gravitating clouds from various random-field approximations for the turbulent gas and then use it to infer the mass distribution of collapsed objects (stars). This means that these theories essentially predict the initial mass function of stars (IMF) from isothermal turbulence and gravity.


Isothermal turbulence with gravity is inherently scale-free\footnote{Scale-free in this context means the equations governing the system's evolution have no inherent scales, but initial conditions can still imprint a mass scale (e.g. Jeans mass). Note that the scale-free statement only applies above the dissipation scale, which is negligible in astrophysical applications.} \citep{McKee_SF_theory,SF_big_problems}, so explaining the turnover in the IMF requires a mass scale from either additional, non-scale-free physics \citep[e.g. protostellar heating, see][]{Krumholz_stellar_mass_origin} or from initial conditions. The specific set of \myquote{turbulent fragmentation} models mentioned here all fall into the latter category: they predict that the initial turbulent properties imprint a mass scale ($M_{\rm sonic}$, the mass of a self-gravitating sphere of gas with transonic turbulence, see Eq. \ref{eq:sonic_mass} later), where the mass distribution starts deviating from the scale-free result (a power-law). However, other works (including some of the same authors) have argued that in a scale-free fragmentation cascade the initial conditions are quickly \myquote{forgotten} by the system \citep{guszejnov_feedback_necessity,Guszejnov_scaling_laws} so the turnover in the IMF can \emph{only} come from additional physics \citep[e.g.][]{larson2005,jappsen2005,bonnell06a}.

There has been significant effort to numerically verify these claims. Most simulations find that \myquote{supersonic clouds} (we will define this rigorously below) fragment into a spectrum of smaller objects \citep[e.g.][]{Goodwin04a,Dobbs06a,Walch12a,Murray_2015_turb_sim} while \myquote{subsonic clouds} undergo monolithic collapse \citep[e.g.][]{VazquezSemadeni_2003,Gong_Ostriker_2009_protostar_form,Gong_Ostriker_2011_conv_flows} similar to the analytical Larson-Penston solution \citep{Larson_1969,Penston_1969}. Of the few convergence studies for the fragmenting case, some report non-convergence up to the highest probed numerical resolution \citep[see e.g.][]{Martel_numerical_sim_convergence,Kratter10a, Hennebelle_Lee_isoT_sim,Federrath_2017_IMF_converge_proceedings}, but a couple have claimed convergence \citep{Gong_Ostriker_2015,Haugbolle_Padoan_isot_IMF} in the mass function (despite the fact that their absolute resolution is comparable or lower than the studies claiming non-convergence).



In at least one special case it is clear that non-convergence is the physically correct answer: \citet{Inutsuka_Miyama_1992} show analytically that isothermal filaments collapse to infinite line density faster than they can fragment along their length, an outcome that cannot be described as collapse to one or more point masses. Thus a simulation of this particular configuration, or of one that evolves into it (e.g., the slowly-rotating Gaussian cloud test of \citealt{Boss_1991_fragmentation}), should not produce a converged outcome for the mass distribution of point-like fragments. Careful resolution studies  confirm this \citep{Boss_2000}: as long as the gas remains isothermal and the simulation resolves the local Jeans length, no fragmentation occurs. Fragmentation only occurs when the rising density of the filament drives the local Jeans length below the maximum allowed resolution, and the resulting fragment masses are determined entirely by the choice of maximum resolution. There is no converged answer. However, it is not clear if this result applies only to the special case of an isothermal filament, or if non-convergence is the typical outcome for isothermal collapse.

Therefore, in this paper we use extremely high-resolution simulations, reaching a maximum density resolution orders of magnitudes higher than the previous studies, to follow the evolution of a self-gravitating isothermal ball of gas, in order to explore the following questions:
\begin{itemize}
\item What are the conditions that determine when a cloud will fragment vs collapse monolithically?
\item Do the initial conditions (e.g. sonic mass, Jeans mass) imprint a mass scale into the mass function of the final fragments?
\item Is there a converged low-mass cut-off for an isothermal fragmentation cascade, or does it proceed \myquote{indefinitely}?
\end{itemize}
Our paper is organized as follows. Sections \ref{sec:isot}+\ref{sec:method} detail the equations solved and the numerical methods. Section \ref{sec:results} shows our results. We  also detail a number of additional numerical tests in Appendix \ref{sec:num_tests}.


\section{Isothermal Collapse}\label{sec:isot}

An isothermal, self-gravitating fluid (well above the dissipation scale) is completely described by the following closed set of equations:
\begin{eqnarray}
\label{eq:mom_eq}
\frac{\partial }{\partial t}\left(\rho \right)+\nabla\cdot\left(\rho\vvector\right)=0,\nonumber \\
\frac{\partial }{\partial t}\left(\rho \vvector\right)+\nabla\cdot\left(\rho\vvector\otimes\vvector\right)=
-\nabla P-\rho\nabla\Phi,
\end{eqnarray}
where $\rho$ and $\vvector$ are the usual fluid density and velocity, while $P=c_s^2\rho$ is the thermal pressure ($c_s=\rm const.$ is the isothermal sound speed) and $\Phi$ is the gravitational potential ($\nabla^2\Phi=4\pi G\rho$, where $G$ is the gravitational constant). By dividing out the characteristic scales of the system (size: $L_0$, density: $\rho_0$ and sound speed: $c_{s}$) we can make these equations dimensionless:
\begin{eqnarray}
\label{eq:mom_eq_dimless}
\frac{\partial }{\partial \tilde{t}}\left(\tilde{\rho}\right)+\tilde{\nabla}\cdot\left(\tilde{\rho}\tilde{\vvector}\right)=0,\nonumber\\
\frac{\partial }{\partial \tilde{t}}\left(\tilde{\rho} \tilde{\vvector}\right)+\tilde{\nabla}\cdot\left(\tilde{\rho}\tilde{\vvector}\otimes\tilde{\vvector}\right)=
-\tilde{\nabla} \tilde{\rho}-\alpha\tilde{\rho}\tilde{\nabla}\tilde{\Phi},
\end{eqnarray}
where $\tilde{t}\equiv t c_s/L_0$, $\tilde{\nabla}\equiv L_0\nabla$ and $\tilde{\Phi}\equiv\frac{\Phi}{G \rho_0 L_0^2}$, while $\alpha\equiv c_s^2/(G \rho_0 L_0^2)$ is the (thermal) virial parameter. It is useful to introduce the Mach number $\mach^2\equiv\frac{1}{3}\langle ||\vvector||^2/c_{s}^2\rangle=\langle ||\tilde{\vvector}||^2\rangle$. By introducing the virial parameter $\alpha$ and the Mach number $\mach$ we normalize the density and velocity fields (e.g. Gaussian velocity distribution, dispersion set by $\mach$).In other words, the dynamics are \emph{entirely} determined by the two dimensionless parameters $\alpha$ and $\mach$, which are fixed by the initial conditions. The only way to imprint scales on the problem is therefore through these ICs.

\subsection{Usual stability measures}\label{sec:measures}

When discussing the stability of an isothermal ball of gas the literature uses a large number of different quantities to characterize these systems. The most common is the \emph{virial parameter}, which is the ratio of two times the energy in random motion over the potential energy\footnote{Note that it is common in the literature to define the virial parameter without thermal energy \citep[e.g. see][]{federrath_sim_2012}. The mapping between the two definitions is $\alpha_{\rm no\,thermal}=\alpha\frac{\mach^2}{1+\mach^2}$, which is close to unity for supersonic clouds. Note that using this alternative definition does not change our results.}. In our case
\be
\label{eq:virial}
\alpha\equiv \frac{2 E_{\rm random, kin}}{E_{\rm pot}}\sim \frac{ 2 M_{\rm cloud}\left(\frac{3}{2} c_s^2+\frac{1}{2}\langle ||\vvector||^2\rangle\right)}{\frac{3}{5} \frac{G M_{\rm cloud}^2}{R_{\rm cloud}}}=
\frac{5 R_{\rm cloud} c_s^2(1+\mach^2)}{G M_{\rm cloud}}.
\ee
We can similarly define the \emph{thermal virial parameter} that only takes thermal motion into account. which leads to
\be
\label{eq:thermal_virial}
\alpha_{\rm thermal}\equiv\frac{2 E_{\rm thermal}}{E_{\rm pot}}\sim \frac{5 R_{\rm cloud} c_s^2}{G M_{\rm cloud}}=\frac{\alpha}{1+\mach^2}.
\ee
Since the behaviour of fluids drastically changes when they become supersonic, another measure is the \emph{infall Mach number}, the characteristic velocity the infalling material would have (relative to the sound speed), if all the potential energy was transferred to infall motion\footnote{The fact that the mode of collapse changes form monolithic collapse to runaway fragmentation changes around $\mach_{\rm infall}\sim 3$ instead of unity (see Sec. \ref{sec:results} and Fig. \ref{fig:fraction_infall_mach}) indicates that only a fraction of the potential energy is transferred to infall motion.}. In our case this yields 
\be 
\label{eq:mach_infall}
\mach_{\rm infall}\equiv\frac{v_{\rm infall}}{c_s}\sim\frac{\sqrt{\frac{1}{3}\frac{E_{\rm pot}}{M_{\rm cloud}}}}{c_s}\sim\sqrt{\frac{G M_{\rm cloud}}{5 R_{\rm cloud} c_s^2}}=\left(\alpha_{\rm thermal}\right)^{-1/2}.
\ee
Since the collapse of such isothermal clouds is mainly precipitated by the Jeans-like instabilities (whose critical masses are dimensionally equivalent to the Jeans mass) another common measure of stability is the \emph{number of Jeans masses in the initial cloud}
\be 
\label{eq:NJ}
N_J\equiv\frac{M_{\rm Jeans}}{M_{\rm cloud}}\sim\left(\frac{3 G M_{\rm cloud}}{4 \pi R_{\rm cloud} c_s^2}\right)^{3/2}=\left(\frac{15}{4\pi}\right)^{3/2}\mach_{\rm infall}^{3}.
\ee
In the case of turbulent fragmentation the initial turbulence has a characteristic mass scale, the sonic mass $M_{\rm sonic}$. To find it let us suppose that the cloud virializes to $\alpha=1$ as energy is transferred from gravity to turbulent motion. One of the characteristic size scale of turbulence is the sonic length $R_{\rm sonic}$. This is where turbulent dispersion becomes supersonic, so
\be 
\label{eq:Rs}
R_{\rm sonic}\equiv R_{\rm cloud}\frac{c_s^2}{\langle ||\vvector_{\rm turb}||^2\rangle}= \frac{R_{\rm cloud}}{\mach^2}\sim \frac{R}{\frac{2GM_{\rm cloud}}{5R_{\rm cloud} c_s^2}-1},
\ee
where we used the supersonic linewidth-size relation ($v_{\rm turb}^2\propto R$)\footnote{Note that in this expression we have already neglected magnetic fields, for the full expression see \cite{federrath_sim_2012}.}.  $M_{\rm sonic}$ is the mass of a self-gravitating ball of gas with $R_{\rm sonic}$ radius (see \citealt{general_turbulent_fragment}), so
\be
\label{eq:sonic_mass}
M_{\rm sonic}\equiv\frac{2\pi^2}{3}\frac{c_s^2 R_{\rm sonic}}{G}.
\ee
With Eqs. \ref{eq:mach_infall}, \ref{eq:Rs} and \ref{eq:sonic_mass} we can formulate the \emph{number of sonic masses in the initial cloud}
\be 
\label{eq:NS}
N_S\equiv\frac{M_{\rm cloud}}{M_{\rm sonic}}\sim\frac{15}{2 \pi^2}\mach_{\rm infall}^2\lbra 2\mach_{\rm infall}^2-1\rbra\approx\frac{15}{2\pi^2}\mach_{\rm infall}^4.
\ee
Note that $\alpha_{\rm thermal}$, $N_J$ and $N_S$ can be all expressed with the infall Mach number (see Eqs. \ref{eq:thermal_virial}-\ref{eq:NS}) so we use only $\mach_{\rm infall}$ as a proxy for all of them for the remainder of this paper.

\section{Simulations}\label{sec:method}

For our simulation we use the GIZMO code (\citealt{Hopkins2015_GIZMO})\footnote{\url{http://www.tapir.caltech.edu/~phopkins/Site/GIZMO.html}}, with the mesh-free Godunov ``MFM'' method for hydrodynamics \citep{Hopkins2015_GIZMO}. Note that we get similar results with other numerical schemes (e.g. SPH), see Appendix \ref{sec:numerics_test}. Self-gravity is included with fully-adaptive force and hydrodynamic resolution - no minimum force length is enforced. Since we are simulating an isothermal system with only self-gravity, the problem is scale-free and we can work in code units of $L=2$, $c_s=1$, $G=1$, where $L$ is the initial size of the box, $c_s$ is the sound speed of the gas and $G$ is the gravitational constant. We start by performing an isothermal driven turbulent box simulation without self-gravity \citep[e.g.][]{schmidt_supersonic_sim,Federrath_sim_compare,Price_2010_turbulent_box} in which the driving force is realized as an Orstein-Uhlenbeck process following \citet{bauerspringel2012}, and consists of a natural mix of compressive and solenoidal modes ($E_{\rm solenoidal}=2 E_{\rm compressive}$). After several crossing times the root-mean-square Mach number saturates to $\mach \sim 1$, and $\tilde{\rho}$ and $\tilde{\vvector}$ are extracted from the simulation to construct the initial conditions of the simulation with self-gravity. These are then rescaled in the following way (using $\alpha$ and $\mach$, the two parameters of isothermal turbulence):
\begin{itemize}
\item Velocities are rescaled so that $\langle||\tilde{\vvector}||^2\rangle=3\mach^2$.
\item The average density $\langle \rho \rangle$ is rescaled to satisfy Eq. \ref{eq:virial} for the specified $\alpha$ virial parameter.
\item The relative density fluctuations are rescaled to satisfy $\langle|\ln\tilde{\rho}|^2\rangle=\ln\left(1+b^2\mach^2\right)$ \citep[see][]{Federrath_turbulence_compressive_PDF}, where $b=1/2$ is the ratio of compressive and solenoidal driving in our initial condition. Effectively this means $\rho=f\left(1-\frac{\langle\rho_{\rm old}\rangle}{\rho_{\rm old}}\right)\langle\rho\rangle+\langle\rho\rangle$, where $\langle\rho\rangle$ is set in the previous step and $f$ is the appropriate scaling factor.
\end{itemize}
Note that in these initial conditions the density and velocity fields are not fully self-consistent. In Appendix \ref{sec:driving} we show that using proper turbulent initialization\footnote{By proper turbulent initialization we mean applying turbulent driving to the system without gravity until statistical equilibrium is reached, then \myquote{turning on} gravity.} does not affect our results. We also show that our results are insensitive to our choice of decaying or driven turbulence during collapse as well as the compressive/solenoidal fraction of the driving\footnote{Note that the insensitivity of the mass function to the initial turbulent fluctuations (in isothermal systems) has already been shown by \cite{Girichidis_2011_isoT_sim}}.

The simulation starts out with $M_{\rm cloud}/\Delta m$ gas particles, where $\Delta m$ is our mass resolution (see Table \ref{tab:resolutions} for details). These particles evolve (now with fully-adaptive self-gravity) following a discretized version of Eq. \ref{eq:mom_eq_dimless} \citep[see][]{Hopkins2015_GIZMO}. They are turned into collapsed objects (sink particles) if they satisfy the following criteria, motivated by \citealt{Federrath_2010_sink_particle}:
\begin{enumerate}
\item They are locally self-gravitating at the resolution scale using the criteria from \citealt{hopkins2013_sf_criterion}.
\item The mean density of this structure exceeds some $\rho_{\rm max}$, at this point the thermal Jeans mass becomes unresolved following the Truelove criterion \citep{truelove_1997_dens_condition}.
\item They are part of a converging flow ($\nabla\cdot\vvector<0$).
\item They are the densest of all particles within the stencil of interacting hydrodynamic cells, and there is no other sink particle within the kernel radius enclosing these interacting cells.
\end{enumerate}
These sink particles can grow by accreting gas from their surroundings if the gas is gravitationally bound to the sink, within a hydrodynamic stencil, and not tightly bound to any other sink particle. In Appendix \ref{sec:sink_test} we explore the effects of our choice of sink particle parameters.

Due to finite resolution our simulation can not resolve the evolution and fragmentation of arbitrarily small structures. This means that we set our mass resolution to the Jeans mass corresponding to $\rho_{\rm max}$ \citep[based on][]{truelove_1997_dens_condition}, so

\begin{eqnarray}
M_{\rm Jeans}(\rho_{\rm max})\sim \Delta m\,\rightarrow\, \rho_{\rm max}\sim \frac{c_s^6}{G^3 \Delta m^2},\nonumber \\
\rho_{\rm max}\propto M_{\rm cloud}(\alpha)^{-2}\left(\frac{\Delta m}{M_{\rm cloud}}\right)^{-2}\propto \lbra \frac{\Delta m}{M_{\rm cloud}} \rbra^{-2} \alpha^2.
\label{eq:rhocrit}
\end{eqnarray}
In this paper we examine the effects of varying the two physical parameters (the virial parameter $\alpha$ and the initial turbulent Mach number $\mach$) on the evolution of an isothermal cloud. To ensure that our results are physical we carry out a resolution study by varying $\frac{\Delta m}{M_{\rm cloud}}$. A number of further tests for numerical effects are also carried out. They are detailed in Appendix \ref{sec:num_tests}. All simulations (with one exception noted) are run until the gas is largely exhausted and the sink particle IMF has remained stable for at least 2 cloud dynamical times.


\section{Results}\label{sec:results}

\begin{figure*}
\begin {center}
\includegraphics[width=0.485\linewidth]{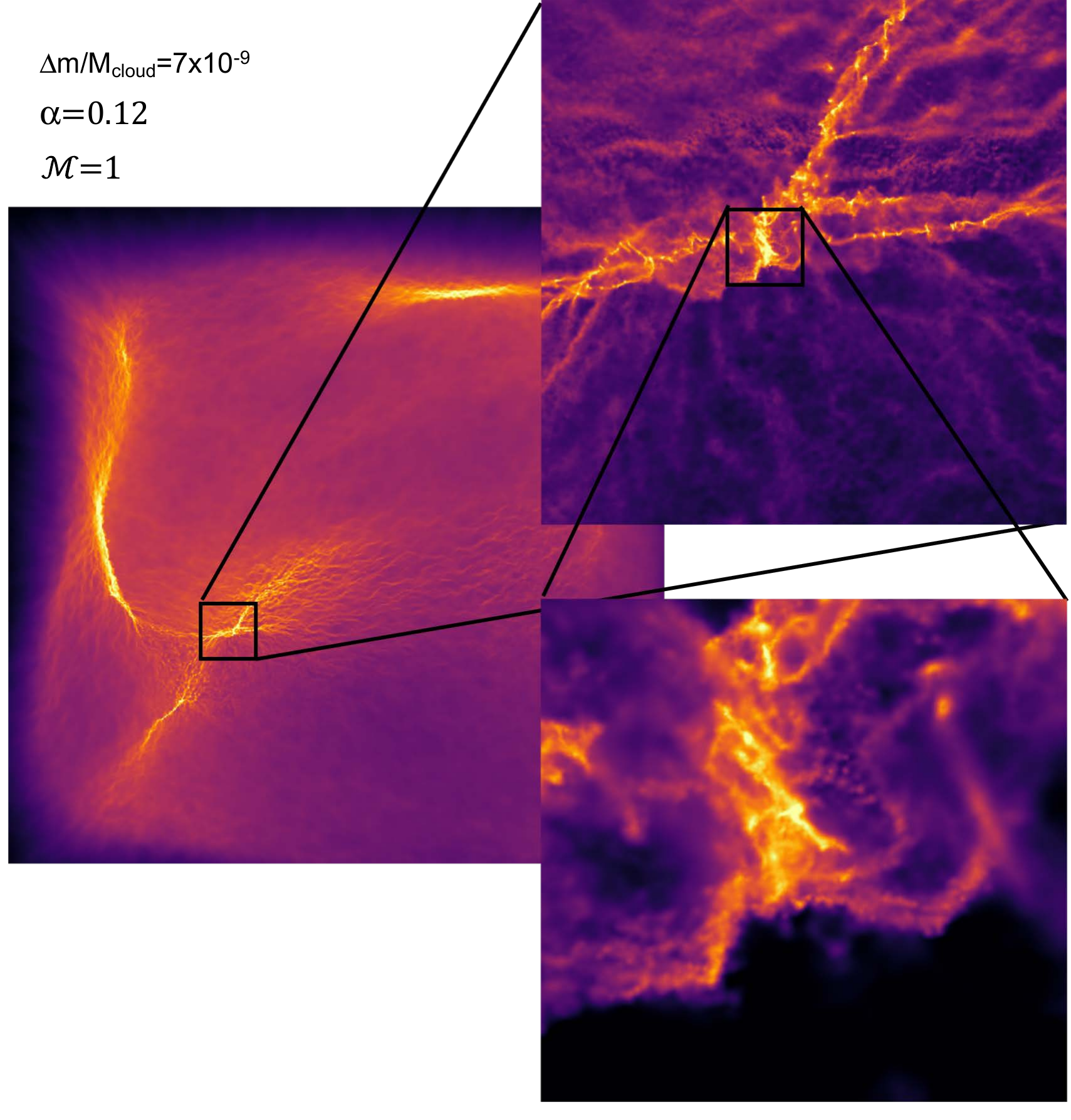}
\includegraphics[width=0.485\linewidth]{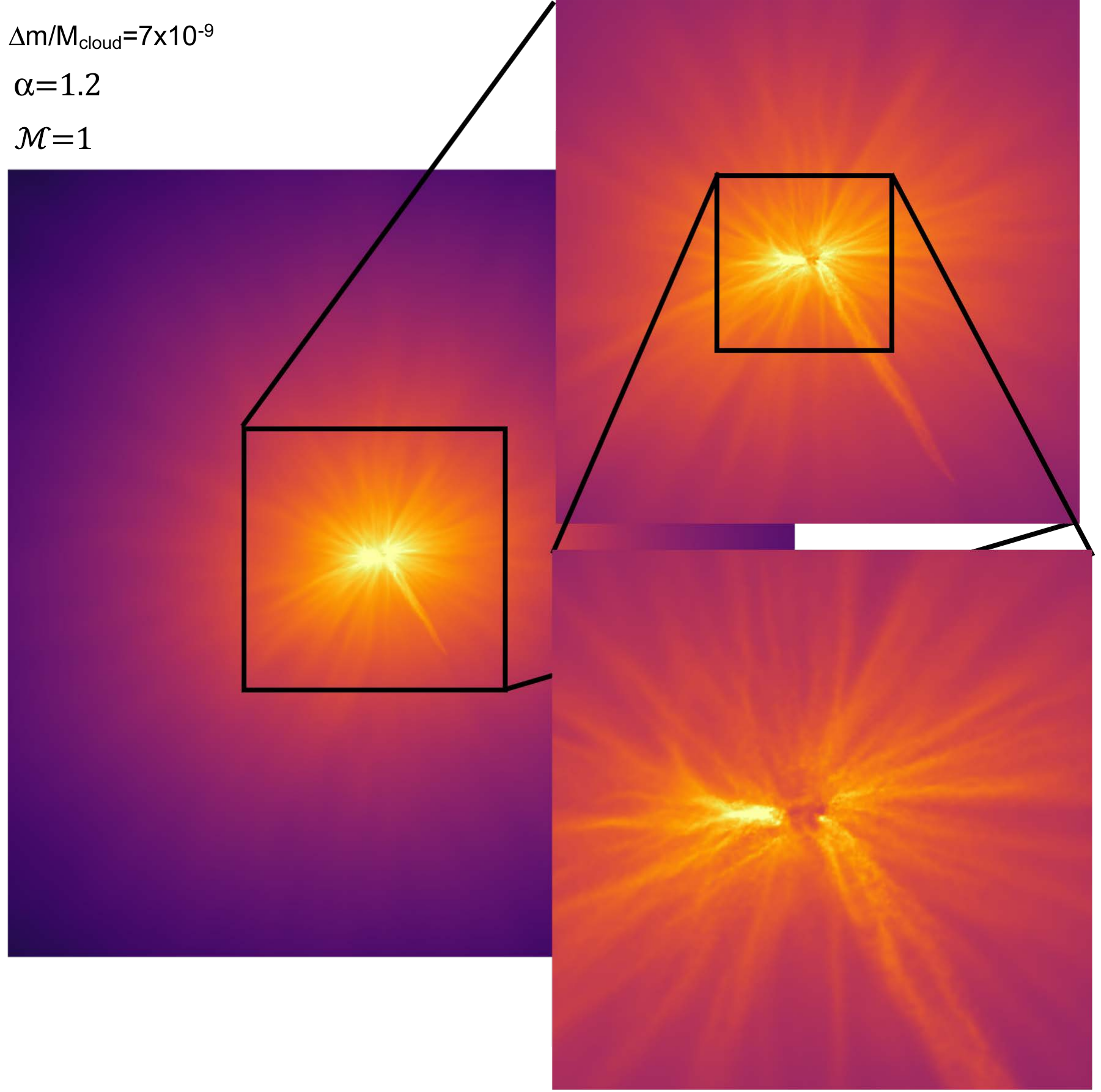}
\caption{Typical density maps for isothermal fragmentation (left) and monolithic collapse (right). On each figure the colormap is stretched over a 2 dex interval. In the fragmenting case (left) supersonic infall creates dense, filamentary structures with high density \myquote{beads} embedded in them. Many of these structures are self-gravitating and undergo gravitational collapse, either forming sink particles or further fragmenting into even smaller objects. In case of monolithic collapse there is only a single high density region at the centre of the cloud, which accretes most of the gas.}
\label{fig:density_maps}
\end {center}
\end{figure*}

\begin{figure*}
\begin {center}
\includegraphics[width=0.45 \linewidth]{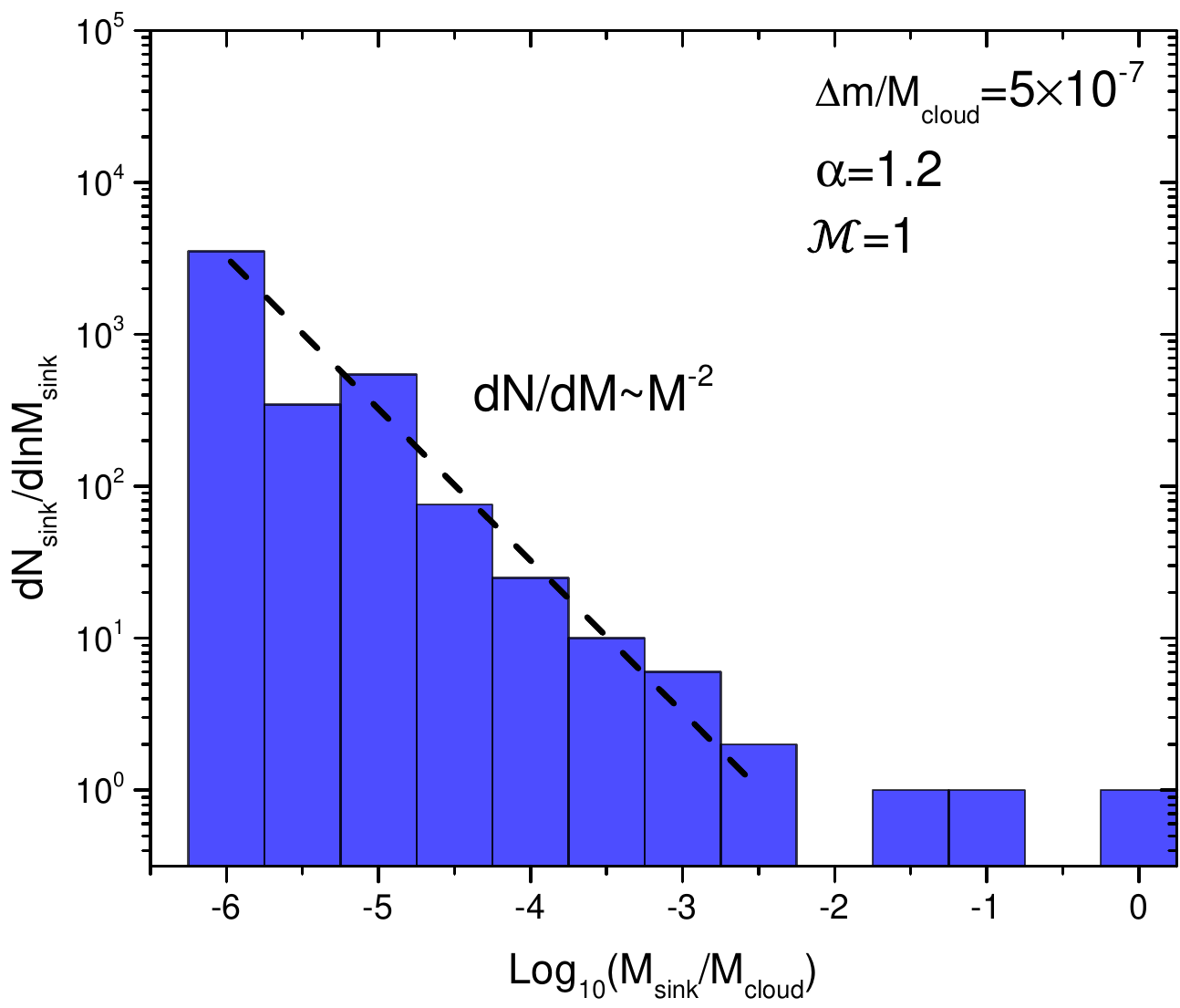}
\includegraphics[width=0.45 \linewidth]{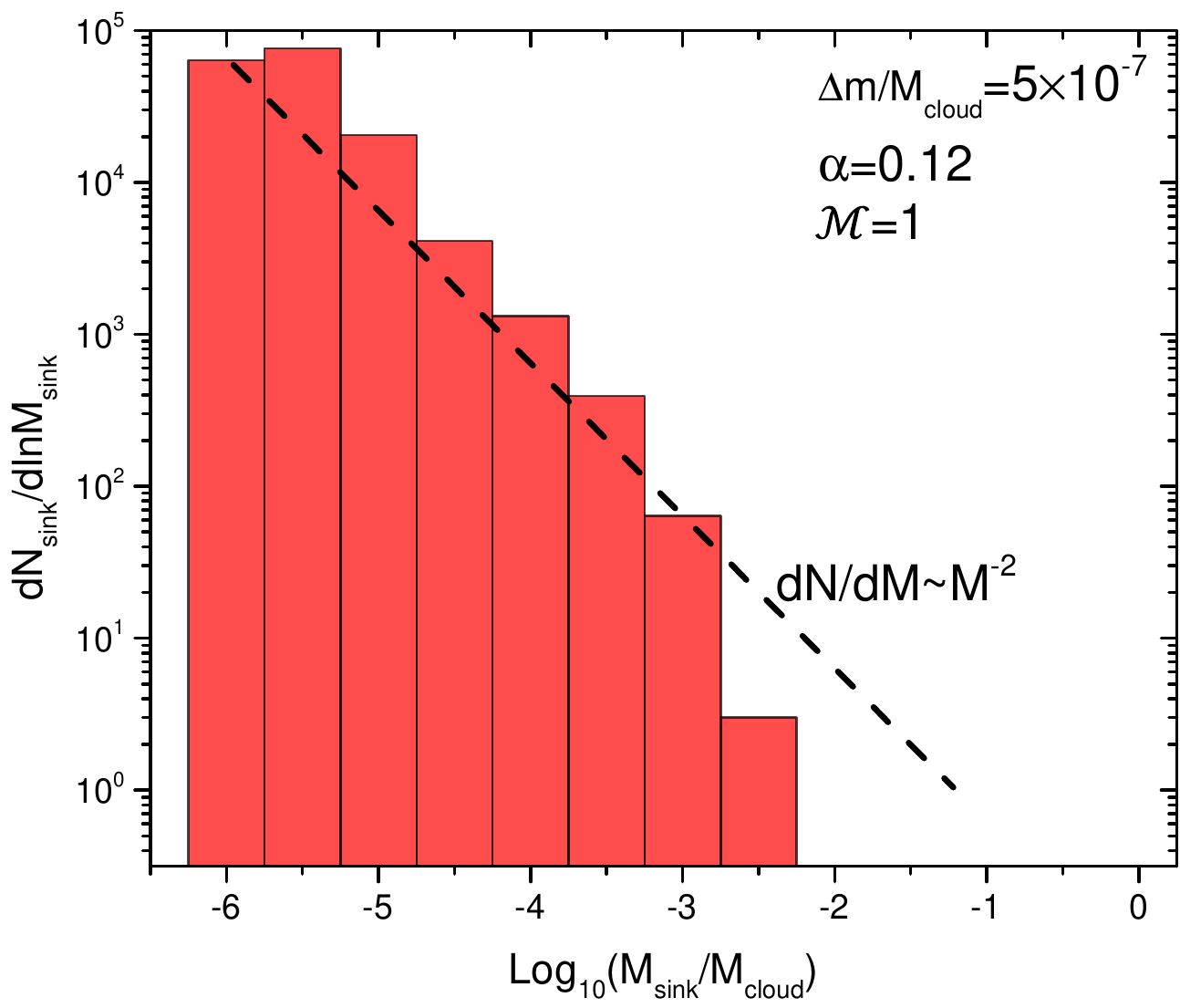}\\
\includegraphics[width=0.45 \linewidth]{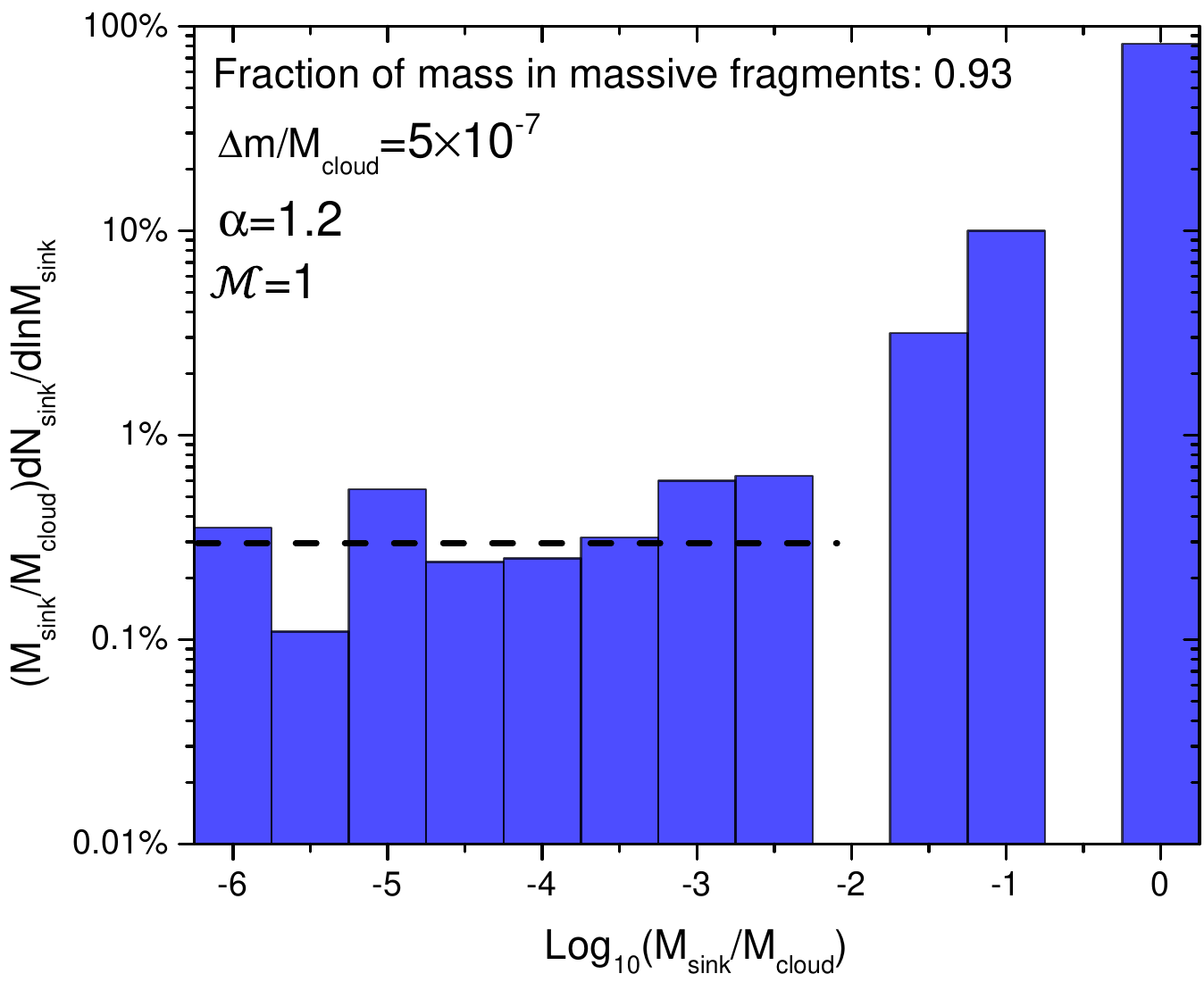}
\includegraphics[width=0.45 \linewidth]{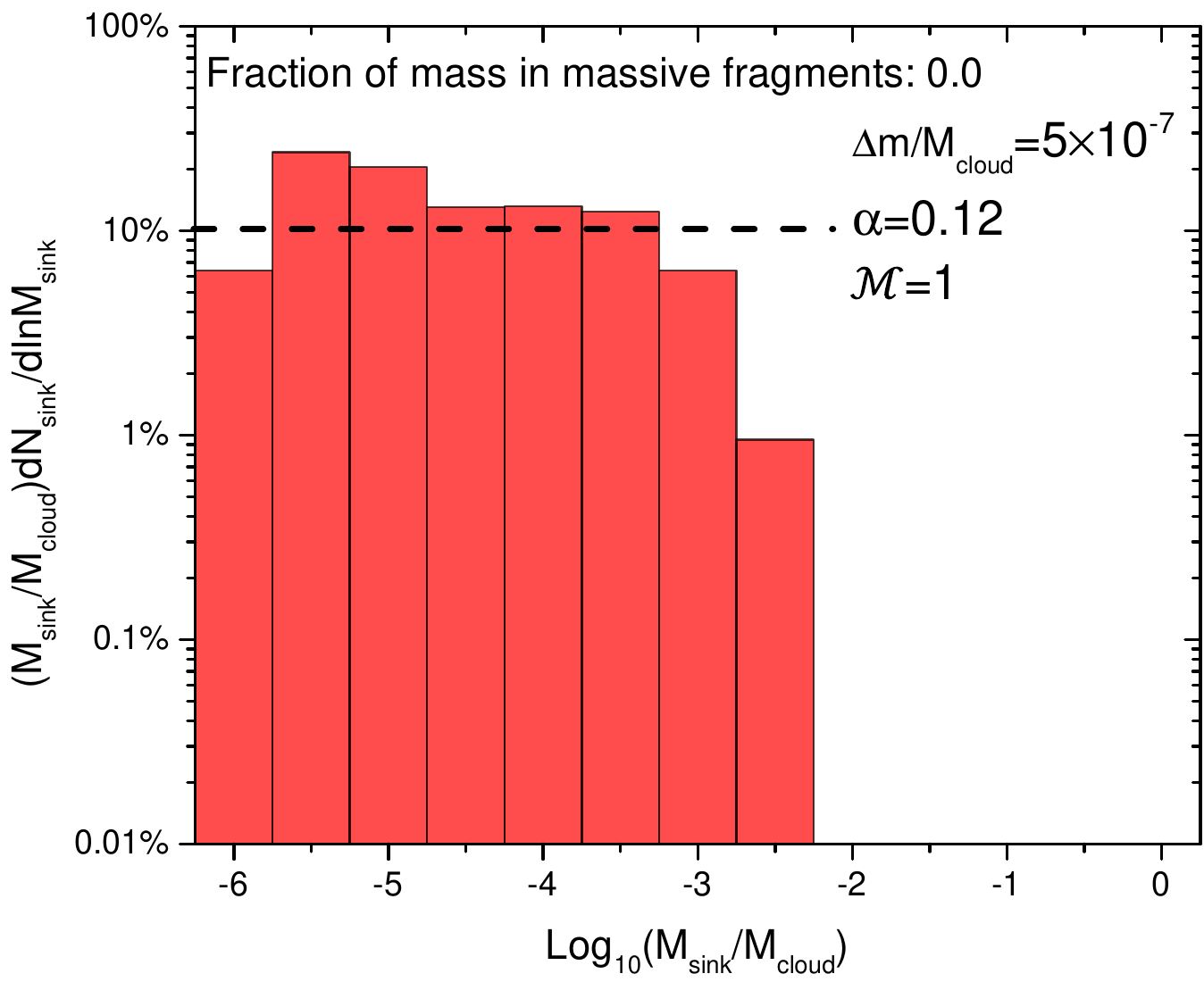}\\
\vspace{-0.2cm}
\caption{The final mass distribution of sink particles in isothermal gravitational collapse for different virial parameters $\alpha$ and fixed initial turbulent Mach number of $\mach=1$, demonstrating the two modes of isothermal collapse. The top row shows the number of particles per mass bin (IMF) while the bottom row shows the total mass of sink particles in each (logarithmic) mass bin. The horizontal axis is normalized by the mass of the initial cloud. In the $\alpha=1.2$ case (left, blue) most of the mass ends up in a single object comparable in mass to the initial cloud (\emph{monolithic collapse}). Meanwhile, in the $\alpha=0.12$ case (right, red) most of the mass ends up in objects with much lower masses than the initial cloud (\emph{fragmentation}). In both cases the low-mass end roughly has equal mass in each logarithmic bin (this means a -2 power-law slope for the IMF), in agreement with theoretical predictions \citep[e.g.][]{Guszejnov_scaling_laws}.}
\label{fig:hist}
\end {center}
\end{figure*}

We carried out a suite of simulations in the $\alpha$-$\mach$ parameter space (our fiducial resolution is $\Delta m/M_{\rm cloud}=4\times 10^{-6}$) and found two distinct modes of evolution (see Fig. \ref{fig:density_maps} for surface density snapshots and Fig. \ref{fig:hist} for statistics). In the first case the collapse is close to monolithic (most of the mass ends up in several massive objects) while in the second case the cloud fragments during collapse, forming a spectrum of low-mass objects (most of the mass in low-mass objects).

The mass spectrum resulting from fragmentation is the well known $\dderiv N/\dderiv M\propto M^{-2}$ distribution (see Fig. \ref{fig:hist}), which means equal mass at each mass scale (see \citealt{Guszejnov_scaling_laws} and references therein). Note that this mass spectrum is present even in the case of monolithic collapse but only a small fraction of the total mass is bound in these low-mass objects.

\begin{figure}
\begin {center}
\includegraphics[width=0.95\linewidth]{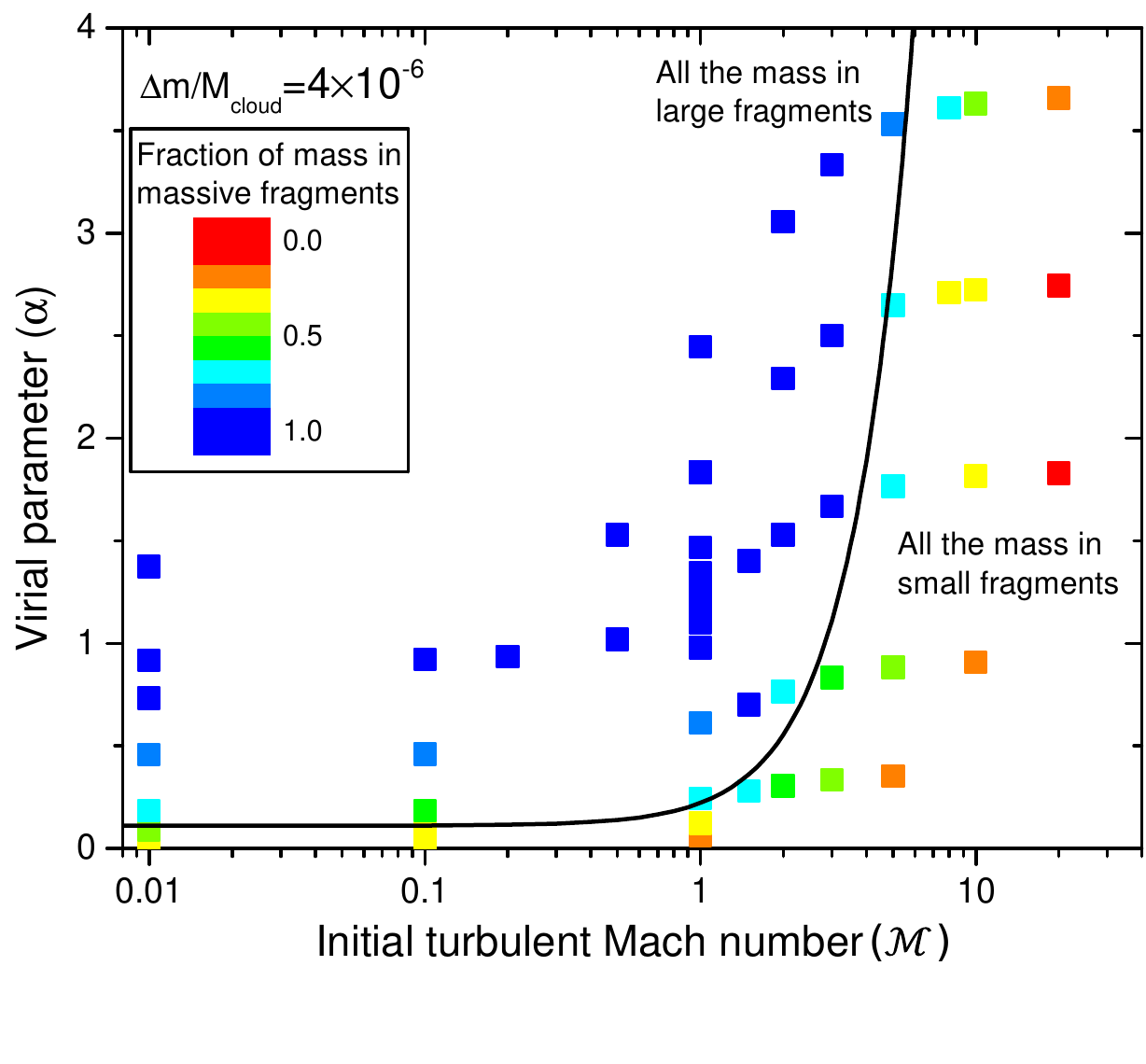}
\vspace{-0.5cm}
\caption{
Fraction of the total cloud mass that ultimately ends up in massive fragments ($M_{\rm sink}>0.1M_{\rm cloud}$) for different initial virial parameters and initial turbulent Mach numbers (blue: most of the mass undergoes monolithic collapse, red: most of the mass ends up in small fragments). It is clear that there is no specific $\alpha$ or initial turbulent $\mach$ value separating the two modes of collapse. However, the boundary is well fitted by $\mach_{\rm infall}=3$  (defined in Eq. \ref{eq:mach_infall}), plotted as a solid black line.}
\label{fig:alpha_mach}
\end {center}
\end{figure}

\begin{figure}
\begin {center}
\includegraphics[width=0.95\linewidth]{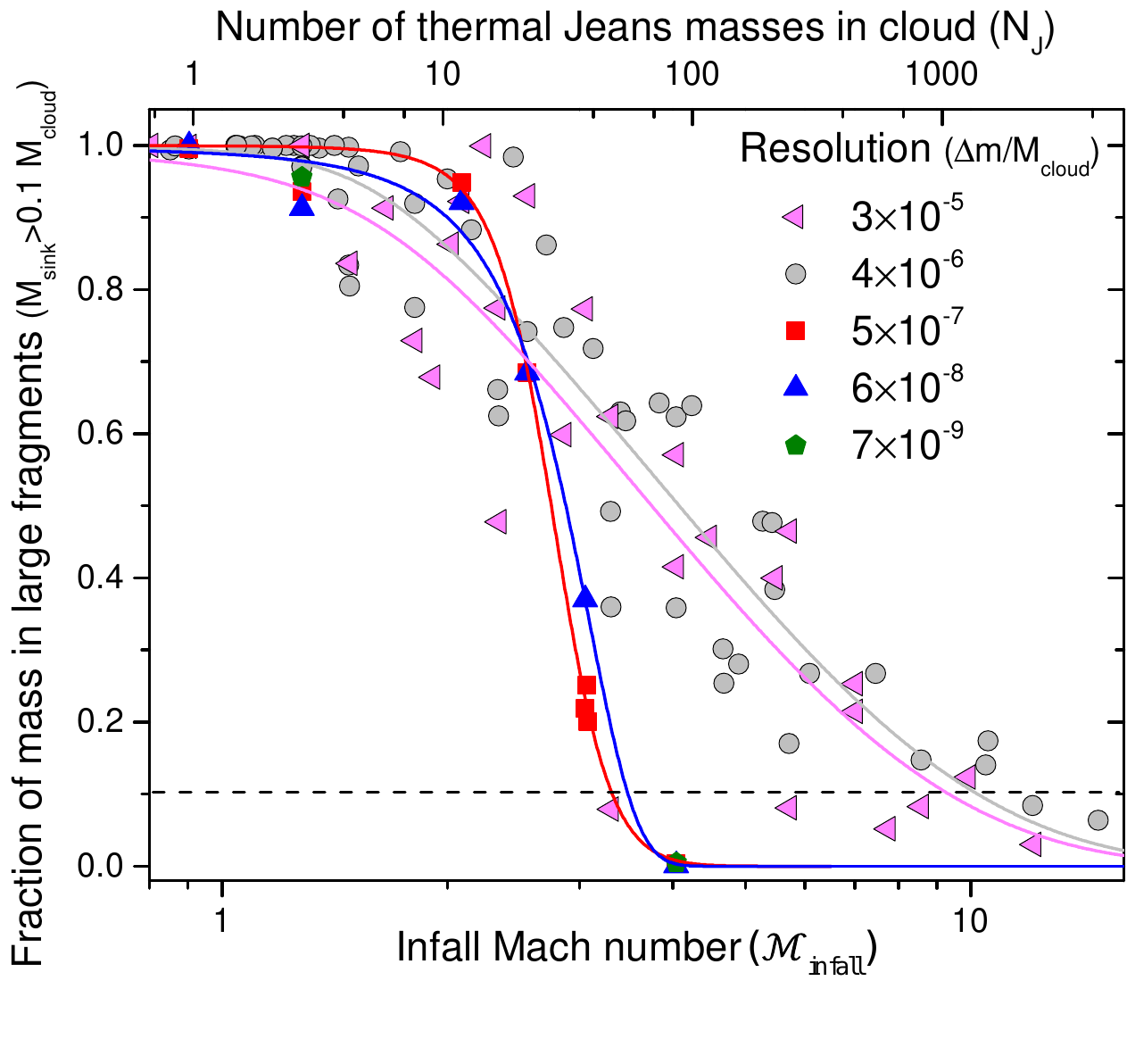}
\vspace{-0.5cm}
\caption{
Fraction of the total cloud mass that ultimately ends up in massive fragments ($M_{\rm sink}>0.1M_{\rm cloud}$) as a function of the infall Mach-number (see Eq. \ref{eq:mach_infall}) or equivalently the number of initial Jeans masses (see Eq. \ref{eq:NJ}). We define massive fragments as having at least 10\% of initial cloud mass. Below this value (dashed line) we plot the mass of the most massive sink particle relative to the cloud. There is a clear transition around $\mach_{\rm infall}\sim 2-4$ between monolithic collapse and fragmentation, we fitted generalized logistic functions $\left(f(x)=\left(1+e^{(x-x_0)/\dderiv x}\right)^{-\nu}\right)$ to the data (solid lines) to make the transition more apparent (no line was fitted at the highest resolution due to the low number of data points, but they lie on the trend line predicted from lower resolution runs). As we go to higher resolutions the transition becomes sharper. Note that the scatter arises from the stochastic nature of the initial conditions (e.g. random velocity field).}
\label{fig:fraction_infall_mach}
\end {center}
\end{figure}

As Fig. \ref{fig:alpha_mach} shows, there is no clear boundary in either the virial parameter $\alpha$ or the Mach number $\mach$ between the two regimes. Instead it is the infall Mach number $\mach_{\rm infall}$ that determines the mode of collapse\footnote{Note that the number of initial Jeans and sonic masses as well as the thermal virial parameter are equally good predictors, because they are all simple functions of $\mach_{\rm infall}$, see Sec. \ref{sec:measures} for how they relate.}. The transition between monolithic collapse and fragmentation occurs around $\mach_{\rm infall}\approx 3$ (see Fig. \ref{fig:fraction_infall_mach}). This boundary roughly corresponds to the point where the characteristic velocity of the infalling material becomes supersonic (this value is >1 because only a fraction of the potential energy is transferred to infall motion, contrary to Eq. \ref{eq:mach_infall}). Considering the filamentary nature of density structures (see Fig \ref{fig:density_maps}), we conjecture that fragmentation is precipitated by localized supersonic infall\footnote{Isothermal supersonic turbulence is has been shown to create filamentary density structures, see e.g. \cite{federrath_sim_2010}. Recent work by \cite{Federrath_2016_filament_universality} has also shown that turbulence is required to reproduce the observed properties of filaments.}. This infall leads to the formation of high density subregions that are self-gravitating and collapse on their own, causing the fragmentation of the cloud. Observation have found a similar trend that a higher $\mach_{\rm infall}$ (or the equivalent $N_J$) leads to higher level of fragmentation within a cloud \citep[e.g.][]{Palau_2015_cloud_fragmentation_Jeans}.

\subsection*{Effect of Resolution on the Mass Distribution}


In the numerical study of isothermal turbulence the dynamic range (resolution) of the simulation plays an important role. If the dynamic range is too small, a multitude of phenomena might not manifest and the results are obscured by artificial edge effects. Since we are primarily interested in the spectrum of self-gravitating objects, let us consider the mass of the smallest resolvable self-gravitating object ($\Delta m$) in a generic simulation of isothermal fragmentation with $N$ particles/grid points. We find that
\begin{itemize}
\item for schemes that follow approximately \emph{uniform mass resolution} (Lagrangian schemes like MFM, SPH, moving mesh methods, and AMR set to ensure equal mass per cell): $\Delta m/M_{\rm cloud}\sim N^{-1}$, trivially.
\item for schemes that follow approximately \emph{uniform spatial resolution} (e.g. uniform Eulerian grids or Lagrangian schemes where the minimum force softening is too large): since there is a spatial resolution $\Delta x$ the smallest resolvable structure has a mass of $\Delta m\sim M_{\rm Jeans}(\Delta x)\sim\frac{c_s^3}{G\rho_{\rm max}}$. Using $\Delta m\sim \rho_{\rm max}\Delta x^3$ we get $\Delta m/M_{\rm cloud}\sim \frac{c_s^2}{G M_{\rm cloud}}\Delta x\propto N^{-1/3}$.
\end{itemize}
This shows that schemes with uniform mass elements (like the Meshless-Finite-Mass scheme we are using) are (as expected by design) inherently superior at resolving mass distributions in Jeans-like collapse for a given number of resolution elements because their low-mass cut-off scales as $N^{-1}$ compared to the $N^{-1/3}$ for uniform spatial resolution schemes (see Table \ref{tab:resolutions} for specifics), provided they use no minimum softening but allow structures to get as dense as needed to reach the Truelove criterion. 

\begin{table*}
	\centering
	\begin{tabular}{ | c | c | c | c | c |c | }
	\hline
	$\Delta m/M_{\rm cloud}$ & $N_{\rm particle}$ & $\Delta x/R_{\rm cloud}\alpha_{\rm thermal}$ & $\rho_{\rm max}/(\rho_0^{\rm cloud} \alpha_{\rm thermal}^3)$ & $\frac{\Delta t_{\rm min}}{t_{\rm dyn, 0}^{\rm cloud}}\alpha_{\rm thermal}^{3/2}$ & $N_{\rm Euler}^{\rm effective}$ \\
	\hline
  $2\times 10^{-4}$ & $(16)^3$ & $9.5\times 10^{-4}$ & $1.2\times 10^{6}$ & $1.8\times 10^{-4}$  & $(4200)^3$ \\
	\hline
  $3\times 10^{-5}$ & $(32)^3$  & $1.2\times 10^{-4}$ & $7.5\times 10^{7}$ & $2.3\times 10^{-5}$  & $(3.3\times 10^4)^3$ \\
	\hline
  $4\times 10^{-6}$ & $(64)^3$  & $1.5\times 10^{-5}$ & $4.8\times 10^{9}$ & $2.9\times 10^{-6}$  & $(2.7\times 10^5)^3$ \\
	\hline
  $5\times 10^{-7}$ & $(128)^3$  & $1.9\times 10^{-6}$ & $3.1\times 10^{11}$ & $3.6\times 10^{-7}$  & $(2.1\times 10^6)^3$ \\
	\hline
  $6\times 10^{-8}$ & $(256)^3$  & $2.3\times 10^{-7}$ & $2.0\times 10^{13}$ & $4.5\times 10^{-8}$  & $(1.7\times 10^7)^3$ \\
  \hline
  $7\times 10^{-9}$ & $(512)^3$  & $2.9\times 10^{-8}$ & $1.3\times 10^{15}$ & $5.6\times 10^{-9}$  & $(1.4\times 10^8)^3$ \\
	\hline
	\end{tabular}
	\caption{Resolution parameters: 1) Fractional mass resolution $\Delta m/M_{\rm cloud}$, 2) Spatial resolution $\Delta x/R_{\rm cloud}\alpha_{\rm thermal}$ where $\rho=\Delta m/\delta x^3$ becomes high enough that the corresponding Jeans mass becomes unresolved ($<\Delta m$), 3) Highest resolvable density $\rho_{\rm max}/(\rho_0^{\rm cloud} \alpha_{\rm thermal}^3)$, the corresponding Jeans mass is $\Delta m$, 4) Smallest resolved time scale $\Delta t_{\rm min}/t_{\rm dyn, 0}^{\rm cloud}\alpha_{\rm thermal}^{3/2}$ where $\Delta t=\frac{\Delta x}{c_s}$ and 5) the number of grid points $N_{\rm Euler}^{\rm effective}$ required in an Eulerian simulation (satisfying $\Delta x=\varphi \Delta x_{\rm grid}$ from \protect\citealt{truelove_1997_dens_condition}, where $\varphi\sim 1$). Note that the CPU cost of these calculations (at best) scale as $N_{\rm particle}\log N_{\rm particle}\times N_{\rm timesteps}$ which means going up one level in resolution (e.g. from $64^3$ to $128^3$) increases the computational cost by roughly a factor of 100.}
\label{tab:resolutions}
\end{table*}

\begin{figure*}
\begin {center}
\includegraphics[width=0.475\linewidth]{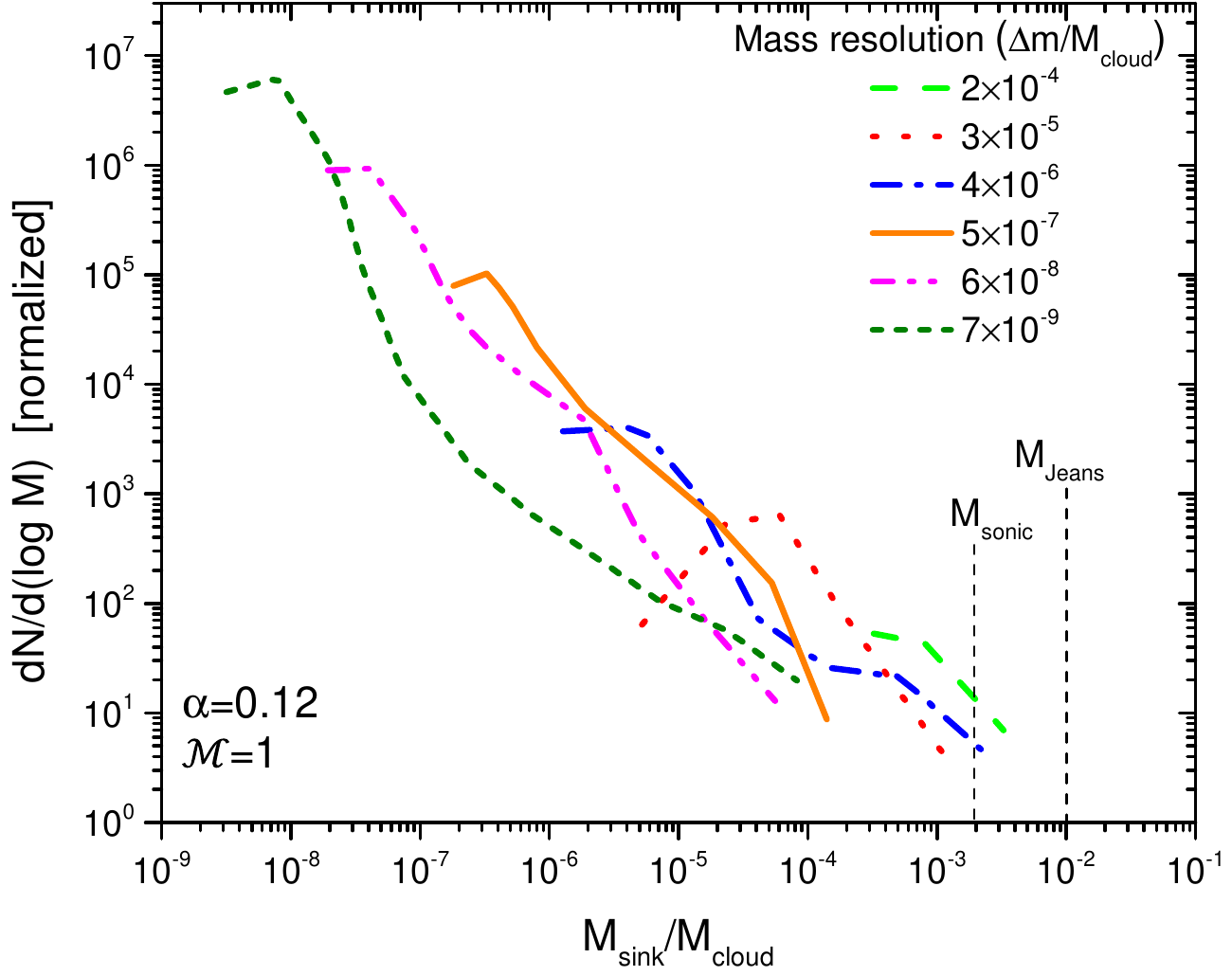}
\includegraphics[width=0.475\linewidth]{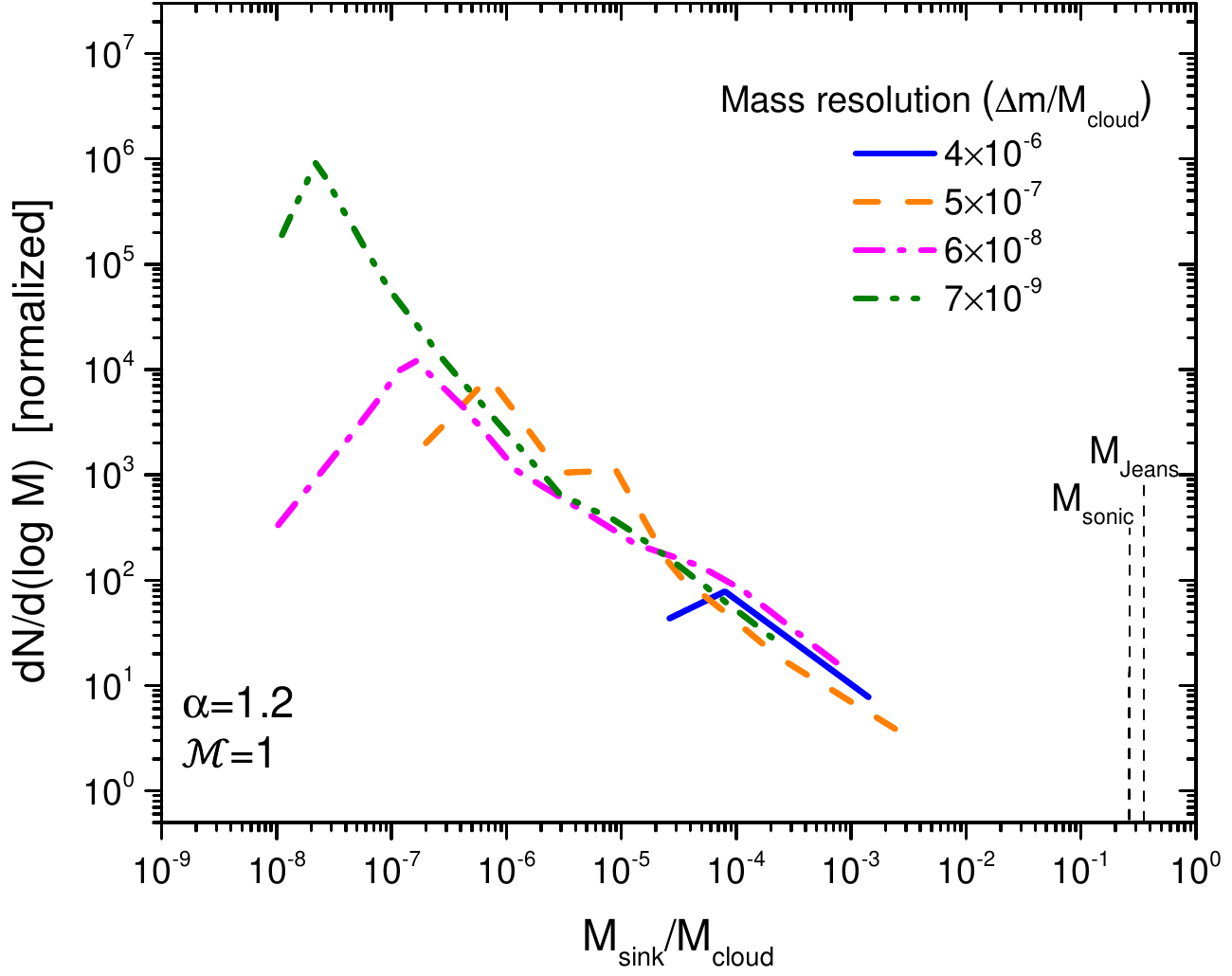}
\caption{The mass distribution of sink particles (IMF) in a fragmenting cloud (Left: $\alpha=0.12$, $\mach=1$, Right: $\alpha=1.2$, $\mach=1$) for different mass resolutions. The dashed lines mark mass scales from initial conditions (sonic mass $M_{\rm sonic}$ and Jeans mass $M_{\rm Jeans}$). For clarity the delta-function-like peaks around unit relative mass were removed from the right figure (see Fig. \ref{fig:hist} for an example). Lower resolution runs are not included in the right figure as they only produced a single sink particle at unit relative mass. It is clear that the peak of the distribution is set by the resolution parameter $\Delta m/M_{\rm cloud}$, initial conditions imprint no scales into the final result. This means that for the infinitely well-resolved case we would get an infinite fragmentation cascade.}
\label{fig:resolution_test}
\end {center}
\end{figure*}

Fig. \ref{fig:resolution_test} shows that the mass distribution in the fragmenting case is close to a power-law with a low-mass cut-off set by the mass resolution of the simulation\footnote{Note that the highest resolution run ($\Delta m/M_{\rm cloud}=7\times10^{-9}$) was not run until completion due to the CPU cost that arises from modelling tightly bound binaries. At this point the system has turned only about 20\% of its mass into sink particles, so we expect the IMF to evolve (e.g. accretion should make it less bottom heavy), but the low-mass cut-off is already established.}. In the monolithic collapse case the distribution of low-mass fragments exhibits a similar behaviour, although the majority of the mass is still contained in several high mass fragments (see Fig. \ref{fig:hist} for reference). This appears to contradict some claims in the literature \citep[e.g.][]{Gong_Ostriker_2015, Haugbolle_Padoan_isot_IMF} that the mass spectrum peak converges around the sonic mass or some other mass scale set by initial conditions. We believe the discrepancy is related to several issues.



First, some authors are using uniform spatial resolution grids (e.g. \citealt{Gong_Ostriker_2015}) for which even the highest resolution calculations can not resolve the fragmentation of substructures due to the unfavourable $\Delta m\propto N^{-1/3}$ scaling\footnote{To reach the resolution of our $\Delta m/M_{\rm cloud}=6\times 10^{-8}$ simulation, $\approx 4\times 10^{21}$ grid points would be needed, far exceeding the capabilities of even large computer clusters.}. Alternatively, it is possible that these simulations start from initial conditions that are reminiscent of the \myquote{monolithic collapse} case (e.g. having substructures in the initial density field that undergo monolithic collapse), and only a small fraction of mass undergoes runaway fragmentation. Although this runaway process is unresolved, the mass function appears converged as most of the mass is bound in objects well above the resolution limit. In some cases authors use adaptive mesh refinement codes and claim convergence, but the data does not support this claim (e.g. see Figs. 4-6 in \citealt{Haugbolle_Padoan_isot_IMF}, where the IMF peak roughly follows the predicted $N^{-1/3}$ trend at higher resolutions\footnote{Although \citet{Haugbolle_Padoan_isot_IMF} use an AMR scheme, their mass resolution follows the unfavourable $\propto N^{-1/3}$ trend, due to the fact that the maximum AMR refinement level is reached before the target mass resolution is.}). Finally, there is substantially greater numerical diffusivity in high $\mathcal{M}$ flows, due to lack of Galilean invariance \citep{springel_hernquist_sph}, which is well-known to generate spurious heating and suppress small-scale structures in the simulation of \myquote{cold} gravitational collapse \citep[see e.g.][]{Hopkins2015_GIZMO}.

\section{Conclusions}\label{sec:consclusions}

We investigated the evolution of self-gravitating, isothermal gas with high-resolution Lagrangian hydrodynamic simulations. We identified two distinct modes of collapse: 
\begin{enumerate}
\item Monolithic collapse (most of the mass ends up in one or a few massive objects)
\item Runaway fragmentation (most of the mass ends up in a spectrum of low-mass fragments, which continues until the resolution limit)
\end{enumerate}
The mode of collapse is set by the infall Mach number $
\mach_{\rm infall}\equiv\frac{v_{\rm infall}}{c_s}\sim\frac{\sqrt{\frac{G M}{R}}}{c_s}$ (equivalent to the initial number of Jeans masses in the cloud), not the initial virial parameter or the Mach number of the initial turbulent dispersion. We conjecture that the difference in behaviour is due to sound waves \myquote{smoothing out} density perturbations when the infall is subsonic leading to a scenario similar to the well-known solutions of isothermal collapse \citep[e.g.][]{Larson_1969,Penston_1969,Shu_1977_isothermal_collapse}, but further tests are needed to verify this claim.

In both modes of collapse we found that the mass distribution of final objects develops a power-law  behaviour at low-masses, close to $\dderiv N/\dderiv M\propto M^{-2}$, in agreement with theoretical expectations \citep[e.g.][]{Elmegreen_1997_imf_fractal_model,Padoan_theory,HC08,Bonnell_2007_competitive_accretion_imf,BallesterosParedes2015}. Note that in the case of monolithic collapse most of the mass is actually in several massive fragments that lie outside this power-law regime but the remaining mass which does not end up in the \myquote{primary} scale sinks forms a power-law distribution, with no lower limit down to the resolution scale.

We conducted a resolution study to examine whether the low-mass cut-off of the power-law in the mass distribution is determined by the initial conditions of the cloud (e.g. its virial parameter or initial turbulent properties) or by mass resolution. We found that there is no convergence in the low-mass spectrum that appears in either mode of collapse. In other words: the fragmentation goes well below the initial Jeans mass, down to the mass resolution. This agrees well with several studies \citep[e.g.][]{Martel_numerical_sim_convergence,Kratter10a,Hennebelle_Lee_isoT_sim,Federrath_2017_IMF_converge_proceedings}. However, these results along with ours do appear to contradict some studies in the literature. We believe the discrepancy is explained by different simulation methods and the much wider dynamic range probed in this study. 


It is a common argument that subsonic structures do not fragment, so the population of such structures (e.g. cores in star formation), whose characteristic mass is set by the large-scale turbulent properties (e.g. sonic mass, see \citealt{HC08,core_imf}), influence the final mass distribution. This is not the case as these structures form in a larger, supersonic cloud that forms supersonic substructures as well. These substructures have different turbulent properties so they spawn a population of subsonic fragments different from their parent. In the end this cascade washes out any effects the initial conditions might have over the low-mass end of the mass spectrum.

We find that once the fragmentation cascade starts, it proceeds to infinitely small scales, similar to the idealized case of \cite{Hoyle_1953}. Initial properties (e.g. virial parameter, turbulent Mach number, Jeans mass, turbulent driving) have no effect on this result, but they may influence the details of the resulting mass distribution (e.g. how close the peak is to the mass resolution). Note that our results only apply to collapsing isothermal gas, additional physics would imprint additional scales, allowing these parameters to exert greater influence on structure formation.

Our results show that an isothermal fragmentation cascade has to be terminated by additional physics (e.g. breakdown of scale-free assumption at high densities); the initial conditions (e.g. sonic mass) imprint no mass scale in the final mass distribution. This means that star formation models that tie the IMF peak to initial turbulent properties \citep[e.g.][]{Padoan_Nordlund_2002_IMF,HC08,excursion_set_ism} need to be modified.

More broadly, these results provide insight into the physical character of isothermal gravito-turbulent fragmentation: it is a self-sustaining process, able to continuously generate enough power in the density field on the smallest scales to drive further fragmentation. The requisite energy to drive these small-scale density perturbations must be produced by {\it local} gravitational collapse, in a manner that is decoupled from energy injection at larger scales \citep[see e.g.][]{Ferrini1983,robertson_adiabatic_heat_fluid,Murray_2015_turb_sim}. This is a very different picture from the classical Kolmogorov energy cascade, in which all kinetic energy originates at large scales and cascades to small scales, with none generated at intermediate scales. Instead we can think of the structure formation process as a mass cascade, where (through fragmentation) mass is transferred to smaller scales \cite[see e.g.][]{Newman_Wasserman_1994_mass_cascade,Field_2008_mass_cascade,Guszejnov_scaling_laws}. Hence self-gravity alters isothermal turbulence in a fundamental way. It follows that any model of the ISM based upon the properties of non-self-gravitating isothermal turbulence will fail to describe the internal dynamics of the self-gravitating objects that form.

\acknowledgments
Support for  PFH, MYG and DG was provided by an Alfred P. Sloan Research Fellowship, NASA ATP Grant NNX14AH35G, and NSF Collaborative Research Grant \#1411920 and CAREER grant \#1455342. Numerical calculations were run on the Caltech compute clusters `Zwicky' (NSF MRI award $\#$ PHY-0960291) and `Wheeler' and allocation TG-AST130039 granted by the Extreme Science and Engineering Discovery Environment (XSEDE) supported by the NSF. Parts of this research were supported by the Australian Research Council Centre of Excellence for All Sky Astrophysics in 3 Dimensions (ASTRO 3D), through project number CE170100013. CF~acknowledges funding provided by the Australian Research Council's Discovery Projects (grants~DP150104329 and~DP170100603), the ANU Futures Scheme, and the Australia-Germany Joint Research Cooperation Scheme (UA-DAAD).

\bibliographystyle{mnras}
\bibliography{bibliography}

\begin{thebibliography}{}
\makeatletter
\relax
\def\mn@urlcharsother{\let\do\@makeother \do\$\do\&\do\#\do\^\do\_\do\%\do\~}
\def\mn@doi{\begingroup\mn@urlcharsother \@ifnextchar [ {\mn@doi@}
  {\mn@doi@[]}}
\def\mn@doi@[#1]#2{\def\@tempa{#1}\ifx\@tempa\@empty \href
  {http://dx.doi.org/#2} {doi:#2}\else \href {http://dx.doi.org/#2} {#1}\fi
  \endgroup}
\def\mn@eprint#1#2{\mn@eprint@#1:#2::\@nil}
\def\mn@eprint@arXiv#1{\href {http://arxiv.org/abs/#1} {{\tt arXiv:#1}}}
\def\mn@eprint@dblp#1{\href {http://dblp.uni-trier.de/rec/bibtex/#1.xml}
  {dblp:#1}}
\def\mn@eprint@#1:#2:#3:#4\@nil{\def\@tempa {#1}\def\@tempb {#2}\def\@tempc
  {#3}\ifx \@tempc \@empty \let \@tempc \@tempb \let \@tempb \@tempa \fi \ifx
  \@tempb \@empty \def\@tempb {arXiv}\fi \@ifundefined
  {mn@eprint@\@tempb}{\@tempb:\@tempc}{\expandafter \expandafter \csname
  mn@eprint@\@tempb\endcsname \expandafter{\@tempc}}}

\bibitem[\protect\citeauthoryear{{Ballesteros-Paredes}, {Hartmann},
  {V{\'a}zquez-Semadeni}, {Heitsch}  \&
  {Zamora-Avil{\'e}s}}{{Ballesteros-Paredes}
  et~al.}{2011}]{grav_turb_dispersion_2011}
{Ballesteros-Paredes} J.,  {Hartmann} L.~W.,  {V{\'a}zquez-Semadeni} E.,
  {Heitsch} F.,   {Zamora-Avil{\'e}s} M.~A.,  2011, \mn@doi [\mnras]
  {10.1111/j.1365-2966.2010.17657.x}, \href
  {http://adsabs.harvard.edu/abs/2011MNRAS.411...65B} {411, 65}

\bibitem[\protect\citeauthoryear{{Ballesteros-Paredes}, {Hartmann},
  {P{\'e}rez-Goytia}  \& {Kuznetsova}}{{Ballesteros-Paredes}
  et~al.}{2015}]{BallesterosParedes2015}
{Ballesteros-Paredes} J.,  {Hartmann} L.~W.,  {P{\'e}rez-Goytia} N.,
  {Kuznetsova} A.,  2015, \mn@doi [\mnras] {10.1093/mnras/stv1285}, \href
  {http://adsabs.harvard.edu/abs/2015MNRAS.452..566B} {452, 566}

\bibitem[\protect\citeauthoryear{{Bauer} \& {Springel}}{{Bauer} \&
  {Springel}}{2012}]{bauerspringel2012}
{Bauer} A.,  {Springel} V.,  2012, \mn@doi [\mnras]
  {10.1111/j.1365-2966.2012.21058.x}, \href
  {http://adsabs.harvard.edu/abs/2012MNRAS.423.2558B} {423, 2558}

\bibitem[\protect\citeauthoryear{{Bonnell}, {Clarke}  \& {Bate}}{{Bonnell}
  et~al.}{2006}]{bonnell06a}
{Bonnell} I.~A.,  {Clarke} C.~J.,   {Bate} M.~R.,  2006, \mnras, 368, 1296

\bibitem[\protect\citeauthoryear{{Bonnell}, {Larson}  \& {Zinnecker}}{{Bonnell}
  et~al.}{2007}]{Bonnell_2007_competitive_accretion_imf}
{Bonnell} I.~A.,  {Larson} R.~B.,   {Zinnecker} H.,  2007, Protostars and
  Planets V, \href {http://adsabs.harvard.edu/abs/2007prpl.conf..149B} {pp
  149--164}

\bibitem[\protect\citeauthoryear{{Boss}}{{Boss}}{1991}]{Boss_1991_fragmentation}
{Boss} A.~P.,  1991, \mn@doi [\nat] {10.1038/351298a0}, \href
  {http://adsabs.harvard.edu/abs/1991Natur.351..298B} {351, 298}

\bibitem[\protect\citeauthoryear{{Boss}, {Fisher}, {Klein}  \& {McKee}}{{Boss}
  et~al.}{2000}]{Boss_2000}
{Boss} A.~P.,  {Fisher} R.~T.,  {Klein} R.~I.,   {McKee} C.~F.,  2000, \mn@doi
  [\apj] {10.1086/308160}, \href
  {http://adsabs.harvard.edu/cgi-bin/nph-bib\_query?bibcode=2000ApJ...528..325B\&db\_key=AST}
  {528, 325}

\bibitem[\protect\citeauthoryear{{Dobbs}, {Bonnell}  \& {Pringle}}{{Dobbs}
  et~al.}{2006}]{Dobbs06a}
{Dobbs} C.~L.,  {Bonnell} I.~A.,   {Pringle} J.~E.,  2006, \mnras, 371, 1663

\bibitem[\protect\citeauthoryear{{Elmegreen}}{{Elmegreen}}{1997}]{Elmegreen_1997_imf_fractal_model}
{Elmegreen} B.~G.,  1997, \mn@doi [\apj] {10.1086/304562}, \href
  {http://adsabs.harvard.edu/abs/1997ApJ...486..944E} {486, 944}

\bibitem[\protect\citeauthoryear{{Federrath}}{{Federrath}}{2016}]{Federrath_2016_filament_universality}
{Federrath} C.,  2016, \mn@doi [\mnras] {10.1093/mnras/stv2880}, \href
  {http://adsabs.harvard.edu/abs/2016MNRAS.457..375F} {457, 375}

\bibitem[\protect\citeauthoryear{{Federrath} \& {Klessen}}{{Federrath} \&
  {Klessen}}{2012}]{federrath_sim_2012}
{Federrath} C.,  {Klessen} R.~S.,  2012, \mn@doi [\apj]
  {10.1088/0004-637X/761/2/156}, \href
  {http://adsabs.harvard.edu/abs/2012ApJ...761..156F} {761, 156}

\bibitem[\protect\citeauthoryear{{Federrath}, {Klessen}  \&
  {Schmidt}}{{Federrath} et~al.}{2008}]{Federrath_turbulence_compressive_PDF}
{Federrath} C.,  {Klessen} R.~S.,   {Schmidt} W.,  2008, \mn@doi [\apjl]
  {10.1086/595280}, \href {http://adsabs.harvard.edu/abs/2008ApJ...688L..79F}
  {688, L79}

\bibitem[\protect\citeauthoryear{{Federrath}, {Roman-Duval}, {Klessen},
  {Schmidt}  \& {Mac Low}}{{Federrath} et~al.}{2010a}]{Federrath_sim_compare}
{Federrath} C.,  {Roman-Duval} J.,  {Klessen} R.~S.,  {Schmidt} W.,   {Mac Low}
  M.-M.,  2010a, \mn@doi [\aap] {10.1051/0004-6361/200912437}, \href
  {http://adsabs.harvard.edu/abs/2010A%26A...512A..81F} {512, A81}

\bibitem[\protect\citeauthoryear{{Federrath}, {Roman-Duval}, {Klessen},
  {Schmidt}  \& {Mac Low}}{{Federrath} et~al.}{2010b}]{federrath_sim_2010}
{Federrath} C.,  {Roman-Duval} J.,  {Klessen} R.~S.,  {Schmidt} W.,   {Mac Low}
  M.-M.,  2010b, \mn@doi [\aap] {10.1051/0004-6361/200912437}, \href
  {http://adsabs.harvard.edu/abs/2010A%26A...512A..81F} {512, A81}

\bibitem[\protect\citeauthoryear{{Federrath}, {Banerjee}, {Clark}  \&
  {Klessen}}{{Federrath} et~al.}{2010c}]{Federrath_2010_sink_particle}
{Federrath} C.,  {Banerjee} R.,  {Clark} P.~C.,   {Klessen} R.~S.,  2010c,
  \mn@doi [\apj] {10.1088/0004-637X/713/1/269}, \href
  {http://adsabs.harvard.edu/abs/2010ApJ...713..269F} {713, 269}

\bibitem[\protect\citeauthoryear{{Federrath} et~al.,}{{Federrath}
  et~al.}{2017a}]{Federrath_2017_turb_driving}
{Federrath} C.,  et~al., 2017a, in {Crocker} R.~M.,  {Longmore} S.~N.,
  {Bicknell} G.~V.,  eds,  IAU Symposium Vol. 322, The Multi-Messenger
  Astrophysics of the Galactic Centre. pp 123--128 (\mn@eprint {arXiv}
  {1609.08726}), \mn@doi{10.1017/S1743921316012357}

\bibitem[\protect\citeauthoryear{{Federrath}, {Krumholz}  \&
  {Hopkins}}{{Federrath}
  et~al.}{2017b}]{Federrath_2017_IMF_converge_proceedings}
{Federrath} C.,  {Krumholz} M.,   {Hopkins} P.~F.,  2017b, in Journal of
  Physics Conference Series. p. 012007, \mn@doi{10.1088/1742-6596/837/1/012007}

\bibitem[\protect\citeauthoryear{{Ferrini}, {Marchesoni}  \&
  {Vulpiani}}{{Ferrini} et~al.}{1983}]{Ferrini1983}
{Ferrini} F.,  {Marchesoni} F.,   {Vulpiani} A.,  1983, \mn@doi [\apss]
  {10.1007/BF00661943}, \href
  {http://adsabs.harvard.edu/abs/1983Ap%26SS..96...83F} {96, 83}

\bibitem[\protect\citeauthoryear{{Field}, {Blackman}  \& {Keto}}{{Field}
  et~al.}{2008}]{Field_2008_mass_cascade}
{Field} G.~B.,  {Blackman} E.~G.,   {Keto} E.~R.,  2008, \mn@doi [\mnras]
  {10.1111/j.1365-2966.2007.12609.x}, \href
  {http://adsabs.harvard.edu/abs/2008MNRAS.385..181F} {385, 181}

\bibitem[\protect\citeauthoryear{{Girichidis}, {Federrath}, {Banerjee}  \&
  {Klessen}}{{Girichidis} et~al.}{2011}]{Girichidis_2011_isoT_sim}
{Girichidis} P.,  {Federrath} C.,  {Banerjee} R.,   {Klessen} R.~S.,  2011,
  \mn@doi [\mnras] {10.1111/j.1365-2966.2011.18348.x}, \href
  {http://esoads.eso.org/abs/2011MNRAS.413.2741G} {413, 2741}

\bibitem[\protect\citeauthoryear{{Gong} \& {Ostriker}}{{Gong} \&
  {Ostriker}}{2009}]{Gong_Ostriker_2009_protostar_form}
{Gong} H.,  {Ostriker} E.~C.,  2009, \mn@doi [\apj]
  {10.1088/0004-637X/699/1/230}, \href
  {http://adsabs.harvard.edu/abs/2009ApJ...699..230G} {699, 230}

\bibitem[\protect\citeauthoryear{{Gong} \& {Ostriker}}{{Gong} \&
  {Ostriker}}{2011}]{Gong_Ostriker_2011_conv_flows}
{Gong} H.,  {Ostriker} E.~C.,  2011, \mn@doi [\apj]
  {10.1088/0004-637X/729/2/120}, \href
  {http://adsabs.harvard.edu/abs/2011ApJ...729..120G} {729, 120}

\bibitem[\protect\citeauthoryear{{Gong} \& {Ostriker}}{{Gong} \&
  {Ostriker}}{2015}]{Gong_Ostriker_2015}
{Gong} M.,  {Ostriker} E.~C.,  2015, \mn@doi [\apj]
  {10.1088/0004-637X/806/1/31}, \href
  {http://adsabs.harvard.edu/abs/2015ApJ...806...31G} {806, 31}

\bibitem[\protect\citeauthoryear{{Goodwin}, {Whitworth}  \&
  {Ward-Thompson}}{{Goodwin} et~al.}{2004}]{Goodwin04a}
{Goodwin} S.~P.,  {Whitworth} A.~P.,   {Ward-Thompson} D.,  2004, \aap, 414,
  633

\bibitem[\protect\citeauthoryear{{Guszejnov}, {Krumholz}  \&
  {Hopkins}}{{Guszejnov} et~al.}{2016}]{guszejnov_feedback_necessity}
{Guszejnov} D.,  {Krumholz} M.~R.,   {Hopkins} P.~F.,  2016, \mn@doi [\mnras]
  {10.1093/mnras/stw315}, \href
  {http://adsabs.harvard.edu/abs/2016MNRAS.458..673G} {458, 673}

\bibitem[\protect\citeauthoryear{{Guszejnov}, {Hopkins}  \&
  {Grudi{\'c}}}{{Guszejnov} et~al.}{2017}]{Guszejnov_scaling_laws}
{Guszejnov} D.,  {Hopkins} P.~F.,   {Grudi{\'c}} M.~Y.,  2017, preprint, \href
  {http://adsabs.harvard.edu/abs/2017arXiv170705799G} {} (\mn@eprint {arXiv}
  {1707.05799})

\bibitem[\protect\citeauthoryear{{Haugb{\o}lle}, {Padoan}  \&
  {Nordlund}}{{Haugb{\o}lle} et~al.}{2017}]{Haugbolle_Padoan_isot_IMF}
{Haugb{\o}lle} T.,  {Padoan} P.,   {Nordlund} A.,  2017, preprint, \href
  {http://adsabs.harvard.edu/abs/2017arXiv170901078H} {} (\mn@eprint {arXiv}
  {1709.01078})

\bibitem[\protect\citeauthoryear{{Hennebelle} \& {Chabrier}}{{Hennebelle} \&
  {Chabrier}}{2008}]{HC08}
{Hennebelle} P.,  {Chabrier} G.,  2008, \mn@doi [\apj] {10.1086/589916}, \href
  {http://adsabs.harvard.edu/abs/2008ApJ...684..395H} {684, 395}

\bibitem[\protect\citeauthoryear{{Hennebelle} \& {Chabrier}}{{Hennebelle} \&
  {Chabrier}}{2009}]{HC2009}
{Hennebelle} P.,  {Chabrier} G.,  2009, \mn@doi [\apj]
  {10.1088/0004-637X/702/2/1428}, \href
  {http://adsabs.harvard.edu/abs/2009ApJ...702.1428H} {702, 1428}

\bibitem[\protect\citeauthoryear{{Hennebelle} \& {Chabrier}}{{Hennebelle} \&
  {Chabrier}}{2013}]{HC_2013}
{Hennebelle} P.,  {Chabrier} G.,  2013, \mn@doi [\apj]
  {10.1088/0004-637X/770/2/150}, \href
  {http://adsabs.harvard.edu/abs/2013ApJ...770..150H} {770, 150}

\bibitem[\protect\citeauthoryear{{Hopkins}}{{Hopkins}}{2012a}]{excursion_set_ism}
{Hopkins} P.~F.,  2012a, \mn@doi [\mnras] {10.1111/j.1365-2966.2012.20730.x},
  \href {http://adsabs.harvard.edu/abs/2012MNRAS.423.2016H} {423, 2016}

\bibitem[\protect\citeauthoryear{{Hopkins}}{{Hopkins}}{2012b}]{core_imf}
{Hopkins} P.~F.,  2012b, \mn@doi [\mnras] {10.1111/j.1365-2966.2012.20731.x},
  \href {http://adsabs.harvard.edu/abs/2012MNRAS.423.2037H} {423, 2037}

\bibitem[\protect\citeauthoryear{{Hopkins}}{{Hopkins}}{2013a}]{Hopkins_SPH_2013}
{Hopkins} P.~F.,  2013a, \mn@doi [\mnras] {10.1093/mnras/sts210}, \href
  {http://adsabs.harvard.edu/abs/2013MNRAS.428.2840H} {428, 2840}

\bibitem[\protect\citeauthoryear{{Hopkins}}{{Hopkins}}{2013b}]{general_turbulent_fragment}
{Hopkins} P.~F.,  2013b, \mn@doi [\mnras] {10.1093/mnras/sts704}, \href
  {http://adsabs.harvard.edu/abs/2013MNRAS.430.1653H} {430, 1653}

\bibitem[\protect\citeauthoryear{{Hopkins}}{{Hopkins}}{2015}]{Hopkins2015_GIZMO}
{Hopkins} P.~F.,  2015, \mn@doi [\mnras] {10.1093/mnras/stv195}, \href
  {http://adsabs.harvard.edu/abs/2015MNRAS.450...53H} {450, 53}

\bibitem[\protect\citeauthoryear{{Hopkins}, {Narayanan}  \& {Murray}}{{Hopkins}
  et~al.}{2013}]{hopkins2013_sf_criterion}
{Hopkins} P.~F.,  {Narayanan} D.,   {Murray} N.,  2013, \mn@doi [\mnras]
  {10.1093/mnras/stt723}, \href
  {http://adsabs.harvard.edu/abs/2013MNRAS.432.2647H} {432, 2647}

\bibitem[\protect\citeauthoryear{{Hoyle}}{{Hoyle}}{1953}]{Hoyle_1953}
{Hoyle} F.,  1953, \mn@doi [\apj] {10.1086/145780}, \href
  {http://adsabs.harvard.edu/abs/1953ApJ...118..513H} {118, 513}

\bibitem[\protect\citeauthoryear{{Hunter}}{{Hunter}}{1962}]{Hunter_1962}
{Hunter} C.,  1962, \mn@doi [\apj] {10.1086/147410}, \href
  {http://adsabs.harvard.edu/abs/1962ApJ...136..594H} {136, 594}

\bibitem[\protect\citeauthoryear{{Hunter}}{{Hunter}}{1964}]{Hunter_1964}
{Hunter} C.,  1964, \mn@doi [\apj] {10.1086/147786}, \href
  {http://adsabs.harvard.edu/abs/1964ApJ...139..570H} {139, 570}

\bibitem[\protect\citeauthoryear{{Ib{\'a}{\~n}ez-Mej{\'{\i}}a}, {Mac Low},
  {Klessen}  \& {Baczynski}}{{Ib{\'a}{\~n}ez-Mej{\'{\i}}a}
  et~al.}{2016}]{grav_vs_SN_turb_driving_2016}
{Ib{\'a}{\~n}ez-Mej{\'{\i}}a} J.~C.,  {Mac Low} M.-M.,  {Klessen} R.~S.,
  {Baczynski} C.,  2016, \mn@doi [\apj] {10.3847/0004-637X/824/1/41}, \href
  {http://adsabs.harvard.edu/abs/2016ApJ...824...41I} {824, 41}

\bibitem[\protect\citeauthoryear{{Inutsuka} \& {Miyama}}{{Inutsuka} \&
  {Miyama}}{1992}]{Inutsuka_Miyama_1992}
{Inutsuka} S.-I.,  {Miyama} S.~M.,  1992, \mn@doi [\apj] {10.1086/171162},
  \href {http://adsabs.harvard.edu/abs/1992ApJ...388..392I} {388, 392}

\bibitem[\protect\citeauthoryear{{Jappsen}, {Klessen}, {Larson}, {Li}  \& {Mac
  Low}}{{Jappsen} et~al.}{2005}]{jappsen2005}
{Jappsen} A.-K.,  {Klessen} R.~S.,  {Larson} R.~B.,  {Li} Y.,   {Mac Low}
  M.-M.,  2005, \mn@doi [\aap] {10.1051/0004-6361:20042178}, \href
  {http://adsabs.harvard.edu/abs/2005A%26A...435..611J} {435, 611}

\bibitem[\protect\citeauthoryear{{Jeans}}{{Jeans}}{1902}]{Jeans_1902}
{Jeans} J.~H.,  1902, \mn@doi [Philosophical Transactions of the Royal Society
  of London Series A] {10.1098/rsta.1902.0012}, \href
  {http://adsabs.harvard.edu/abs/1902RSPTA.199....1J} {199, 1}

\bibitem[\protect\citeauthoryear{{Kratter}, {Matzner}, {Krumholz}  \&
  {Klein}}{{Kratter} et~al.}{2010}]{Kratter10a}
{Kratter} K.~M.,  {Matzner} C.~D.,  {Krumholz} M.~R.,   {Klein} R.~I.,  2010,
  \mn@doi [\apj] {10.1088/0004-637X/708/2/1585}, \href
  {http://adsabs.harvard.edu/abs/2010ApJ...708.1585K} {708, 1585}

\bibitem[\protect\citeauthoryear{{Krumholz}}{{Krumholz}}{2011}]{Krumholz_stellar_mass_origin}
{Krumholz} M.~R.,  2011, \mn@doi [\apj] {10.1088/0004-637X/743/2/110}, \href
  {http://adsabs.harvard.edu/abs/2011ApJ...743..110K} {743, 110}

\bibitem[\protect\citeauthoryear{{Krumholz}}{{Krumholz}}{2014}]{SF_big_problems}
{Krumholz} M.~R.,  2014, \mn@doi [\physrep] {10.1016/j.physrep.2014.02.001},
  \href {http://adsabs.harvard.edu/abs/2014PhR...539...49K} {539, 49}

\bibitem[\protect\citeauthoryear{{Larson}}{{Larson}}{1969}]{Larson_1969}
{Larson} R.~B.,  1969, \mn@doi [\mnras] {10.1093/mnras/145.3.271}, \href
  {http://adsabs.harvard.edu/abs/1969MNRAS.145..271L} {145, 271}

\bibitem[\protect\citeauthoryear{{Larson}}{{Larson}}{2005}]{larson2005}
{Larson} R.~B.,  2005, \mn@doi [\mnras] {10.1111/j.1365-2966.2005.08881.x},
  \href {http://adsabs.harvard.edu/abs/2005MNRAS.359..211L} {359, 211}

\bibitem[\protect\citeauthoryear{{Lee} \& {Hennebelle}}{{Lee} \&
  {Hennebelle}}{2017}]{Hennebelle_Lee_isoT_sim}
{Lee} Y.-N.,  {Hennebelle} P.,  2017, preprint, \href
  {http://adsabs.harvard.edu/abs/2017arXiv171100319L} {} (\mn@eprint {arXiv}
  {1711.00319})

\bibitem[\protect\citeauthoryear{{Martel}, {Evans}  \& {Shapiro}}{{Martel}
  et~al.}{2006}]{Martel_numerical_sim_convergence}
{Martel} H.,  {Evans} II N.~J.,   {Shapiro} P.~R.,  2006, \mn@doi [\apjs]
  {10.1086/500090}, \href {http://adsabs.harvard.edu/abs/2006ApJS..163..122M}
  {163, 122}

\bibitem[\protect\citeauthoryear{{McKee} \& {Ostriker}}{{McKee} \&
  {Ostriker}}{2007}]{McKee_SF_theory}
{McKee} C.~F.,  {Ostriker} E.~C.,  2007, \mn@doi [\araa]
  {10.1146/annurev.astro.45.051806.110602}, \href
  {http://adsabs.harvard.edu/abs/2007ARA%26A..45..565M} {45, 565}

\bibitem[\protect\citeauthoryear{{Murray}, {Chang}, {Murray}  \&
  {Pittman}}{{Murray} et~al.}{2017}]{Murray_2015_turb_sim}
{Murray} D.~W.,  {Chang} P.,  {Murray} N.~W.,   {Pittman} J.,  2017, \mn@doi
  [\mnras] {10.1093/mnras/stw2796}, \href
  {http://adsabs.harvard.edu/abs/2017MNRAS.465.1316M} {465, 1316}

\bibitem[\protect\citeauthoryear{{Newman} \& {Wasserman}}{{Newman} \&
  {Wasserman}}{1990}]{Newman_Wasserman_1994_mass_cascade}
{Newman} W.~I.,  {Wasserman} I.,  1990, \mn@doi [\apj] {10.1086/168703}, \href
  {http://adsabs.harvard.edu/abs/1990ApJ...354..411N} {354, 411}

\bibitem[\protect\citeauthoryear{{Padoan} \& {Nordlund}}{{Padoan} \&
  {Nordlund}}{2002}]{Padoan_Nordlund_2002_IMF}
{Padoan} P.,  {Nordlund} {\AA}.,  2002, \mn@doi [\apj] {10.1086/341790}, \href
  {http://adsabs.harvard.edu/abs/2002ApJ...576..870P} {576, 870}

\bibitem[\protect\citeauthoryear{{Padoan}, {Nordlund}  \& {Jones}}{{Padoan}
  et~al.}{1997}]{Padoan_theory}
{Padoan} P.,  {Nordlund} A.,   {Jones} B.~J.~T.,  1997, \mnras, \href
  {http://adsabs.harvard.edu/abs/1997MNRAS.288..145P} {288, 145}

\bibitem[\protect\citeauthoryear{{Palau} et~al.,}{{Palau}
  et~al.}{2015}]{Palau_2015_cloud_fragmentation_Jeans}
{Palau} A.,  et~al., 2015, \mn@doi [\mnras] {10.1093/mnras/stv1834}, \href
  {http://adsabs.harvard.edu/abs/2015MNRAS.453.3785P} {453, 3785}

\bibitem[\protect\citeauthoryear{{Penston}}{{Penston}}{1969}]{Penston_1969}
{Penston} M.~V.,  1969, \mn@doi [\mnras] {10.1093/mnras/144.4.425}, \href
  {http://adsabs.harvard.edu/abs/1969MNRAS.144..425P} {144, 425}

\bibitem[\protect\citeauthoryear{{Price} \& {Federrath}}{{Price} \&
  {Federrath}}{2010}]{Price_2010_turbulent_box}
{Price} D.~J.,  {Federrath} C.,  2010, \mn@doi [\mnras]
  {10.1111/j.1365-2966.2010.16810.x}, \href
  {http://adsabs.harvard.edu/abs/2010MNRAS.406.1659P} {406, 1659}

\bibitem[\protect\citeauthoryear{{Robertson} \& {Goldreich}}{{Robertson} \&
  {Goldreich}}{2012}]{robertson_adiabatic_heat_fluid}
{Robertson} B.,  {Goldreich} P.,  2012, \mn@doi [\apjl]
  {10.1088/2041-8205/750/2/L31}, \href
  {http://adsabs.harvard.edu/abs/2012ApJ...750L..31R} {750, L31}

\bibitem[\protect\citeauthoryear{{Schmidt}, {Federrath}, {Hupp}, {Kern}  \&
  {Niemeyer}}{{Schmidt} et~al.}{2009}]{schmidt_supersonic_sim}
{Schmidt} W.,  {Federrath} C.,  {Hupp} M.,  {Kern} S.,   {Niemeyer} J.~C.,
  2009, \mn@doi [\aap] {10.1051/0004-6361:200809967}, \href
  {http://adsabs.harvard.edu/abs/2009A%26A...494..127S} {494, 127}

\bibitem[\protect\citeauthoryear{{Shu}}{{Shu}}{1977}]{Shu_1977_isothermal_collapse}
{Shu} F.~H.,  1977, \mn@doi [\apj] {10.1086/155274}, \href
  {http://adsabs.harvard.edu/abs/1977ApJ...214..488S} {214, 488}

\bibitem[\protect\citeauthoryear{{Springel} \& {Hernquist}}{{Springel} \&
  {Hernquist}}{2002}]{springel_hernquist_sph}
{Springel} V.,  {Hernquist} L.,  2002, \mn@doi [\mnras]
  {10.1046/j.1365-8711.2002.05445.x}, \href
  {http://adsabs.harvard.edu/abs/2002MNRAS.333..649S} {333, 649}

\bibitem[\protect\citeauthoryear{{Tohline}}{{Tohline}}{1980}]{Tohline_1980}
{Tohline} J.~E.,  1980, \mn@doi [\apj] {10.1086/157688}, \href
  {http://adsabs.harvard.edu/abs/1980ApJ...235..866T} {235, 866}

\bibitem[\protect\citeauthoryear{{Truelove}, {Klein}, {McKee}, {Holliman},
  {Howell}  \& {Greenough}}{{Truelove}
  et~al.}{1997}]{truelove_1997_dens_condition}
{Truelove} J.~K.,  {Klein} R.~I.,  {McKee} C.~F.,  {Holliman} II J.~H.,
  {Howell} L.~H.,   {Greenough} J.~A.,  1997, \mn@doi [\apjl] {10.1086/310975},
  \href {http://adsabs.harvard.edu/abs/1997ApJ...489L.179T} {489, L179}

\bibitem[\protect\citeauthoryear{{V{\'a}zquez-Semadeni}, {Ballesteros-Paredes}
  \& {Klessen}}{{V{\'a}zquez-Semadeni} et~al.}{2003}]{VazquezSemadeni_2003}
{V{\'a}zquez-Semadeni} E.,  {Ballesteros-Paredes} J.,   {Klessen} R.~S.,  2003,
  \mn@doi [\apjl] {10.1086/374325}, \href
  {http://adsabs.harvard.edu/abs/2003ApJ...585L.131V} {585, L131}

\bibitem[\protect\citeauthoryear{{Walch}, {Whitworth}  \& {Girichidis}}{{Walch}
  et~al.}{2012}]{Walch12a}
{Walch} S.,  {Whitworth} A.~P.,   {Girichidis} P.,  2012, \mnras, 419, 760

\makeatother
\end{thebibliography}

\appendix

\section{Additional Numerical Tests}\label{sec:num_tests}

\subsection{Effects of perturbed initial conditions}\label{sec:perturb}

\begin{figure}
\begin {center}
\includegraphics[width=0.95\linewidth]{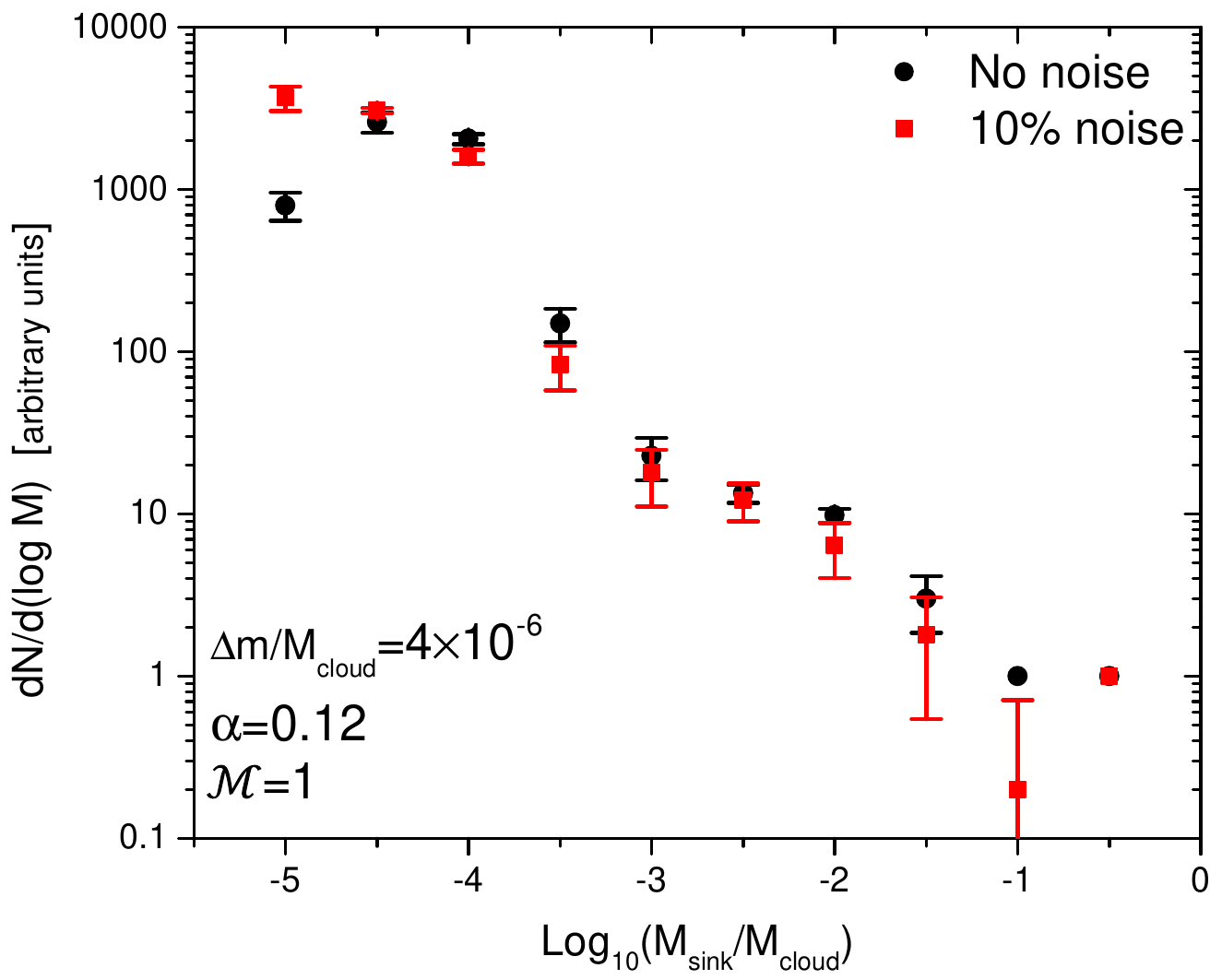}
\caption{The mass distribution of sink particles (IMF) for 5 different simulations of clouds, each with the same initial $\alpha=0.06$ and turbulent Mach number of $\mach=1$, but the ICs are drawn randomly from different times in a turbulent boy simulation (in other words, these are different \myquote{realizations} of the ICs). Points show the median and error bars the 80\% inclusion interval of the sink particle IMF for different realizations. In a second set of simulations we added a random Gaussian perturbation to the initial position and velocity of the gas particles. The IMF shape is is qualitatively consistent for different realizations even if we add significant perturbations onto it.}
\label{fig:perturb}
\end {center}
\end{figure}

Due to the resource intense nature of the simulation, only one initial realization of the initial conditions (e.g. the specific density field) was simulated for a given resolution, virial parameter $\alpha$ and Mach number $\mach$ in Fig. \ref{fig:alpha_mach}. To test for the magnitude of stochastic effects in different realizations (since the system is chaotic) we consider an experiment where we follow the evolution of 5 different random realizations  with the same global Mach number and virial parameter. We also included a set where we added Gaussian noise to the position and velocity of the initial gas particles. Fig. \ref{fig:perturb} shows that the mass distribution of sink particles (IMF) is qualitatively unchanged by these experiments.

\subsection{Effects of turbulent driving}\label{sec:driving}

The simulations mentioned in the main text include no external driving for turbulence as this simplifies the problem and decreases the number of degrees of freedom. Nevertheless, turbulent driving could play an important role in star formation \citep[see e.g.][]{Federrath_2017_turb_driving}, although some authors have argued that turbulence in clouds is driven primarily by their own self-gravity, not an external cascade in the isothermal regime we are focusing on \citep{grav_turb_dispersion_2011,robertson_adiabatic_heat_fluid,grav_vs_SN_turb_driving_2016,Murray_2015_turb_sim}.

\begin{figure}
\begin {center}
\includegraphics[width=0.95\linewidth]{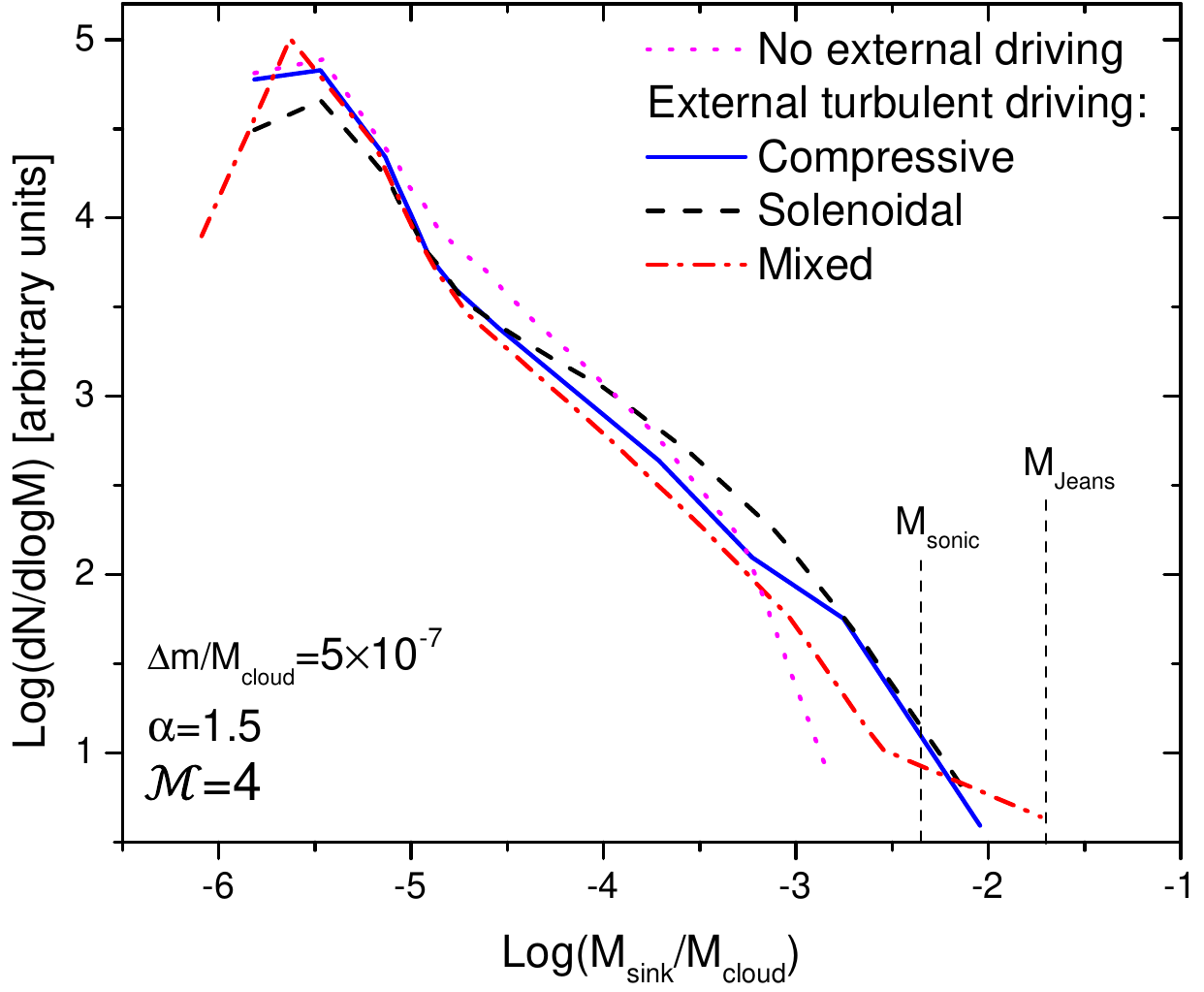}
\caption{The mass distribution of sink particles (IMF) for a cloud with initial $\alpha=0.5$ and $\mach=4$. We compare our fiducial case of non-driven (but still self-gravitating) turbulence with scenarios with different types of turbulent driving. It is clear that the external driving has little to no effect on the final distribution, regardless of the driving method.}
\label{fig:driving}
\end {center}
\end{figure}

To investigate the effects of turbulent driving we carried out several simulations where the initial conditions are generated by driving the turbulence for several dynamical times without self-gravity, then turning on gravity (as in e.g. \citealt{schmidt_supersonic_sim}). Note that unlike the simulations in the main text in these cases the density and velocity fields in the initial conditions are self-consistent with the driving and initial Mach number. Fig. \ref{fig:driving} shows that turbulent driving has no qualitative effects on the resulting IMF.

\subsection{Effects of the hydrodynamic solver}\label{sec:numerics_test}

\begin{figure}
\begin {center}
\includegraphics[width=0.95\linewidth]{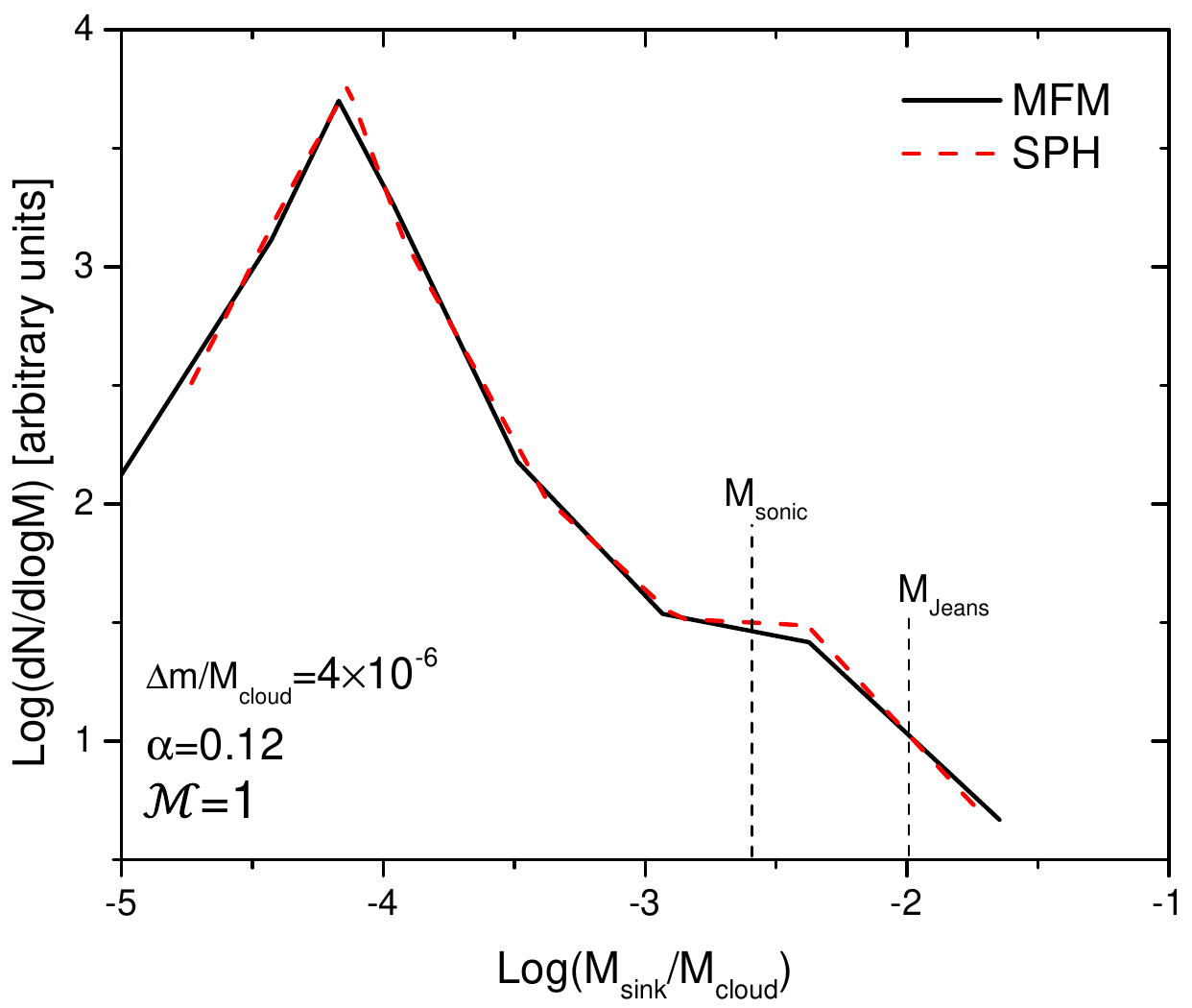}
\caption{The mass distribution of sink particles (IMF) using Meshless Finite-Mass (MFM) and Smoothed-Particle Hydrodynamics (SPH) schemes. The predicted IMF is independent of the details of the hydrodynamics method.}
\label{fig:methods}
\end {center}
\end{figure}

As GIZMO is an inherently multi-method code, we can re-run several simulations with different hydrodynamics schemes, but otherwise identical physics. We compare:
\begin{itemize}
\item The Meshless Finite-Mass (MFM) method \citep{Hopkins2015_GIZMO}, a Lagrangian, finite volume, second order, Godunov method (our default in the text).
\item The \myquote{Pressure-Energy} formulation of smoothed particle hydrodynamics (SPH) \citep{Hopkins_SPH_2013}, which has various improvements over the original GADGET \myquote{Density-Entropy} formulation it is derived from \citep{springel_hernquist_sph}.
\end{itemize}
Although both are Lagrangian methods, the two work quite differently. In MFM, inter-cell fluxes are the obtained by solving a Riemann problem across each effective face between neighbouring cells in such a way that mass fluxes cancel and the cells are moved with the local fluid velocity. In SPH, effective forces between interacting neighbour particles are derived from a discrete particle Lagrangian involving the local fluid properties reconstructed from a kernel density estimator. Despite these differences, we found that our choice of hydro solver these has no qualitative effect on our results (see Fig. \ref{fig:methods}).

\subsection{Effects of the sink particle scheme}\label{sec:sink_test}

\begin{figure}
\begin {center}
\includegraphics[width=0.95\linewidth]{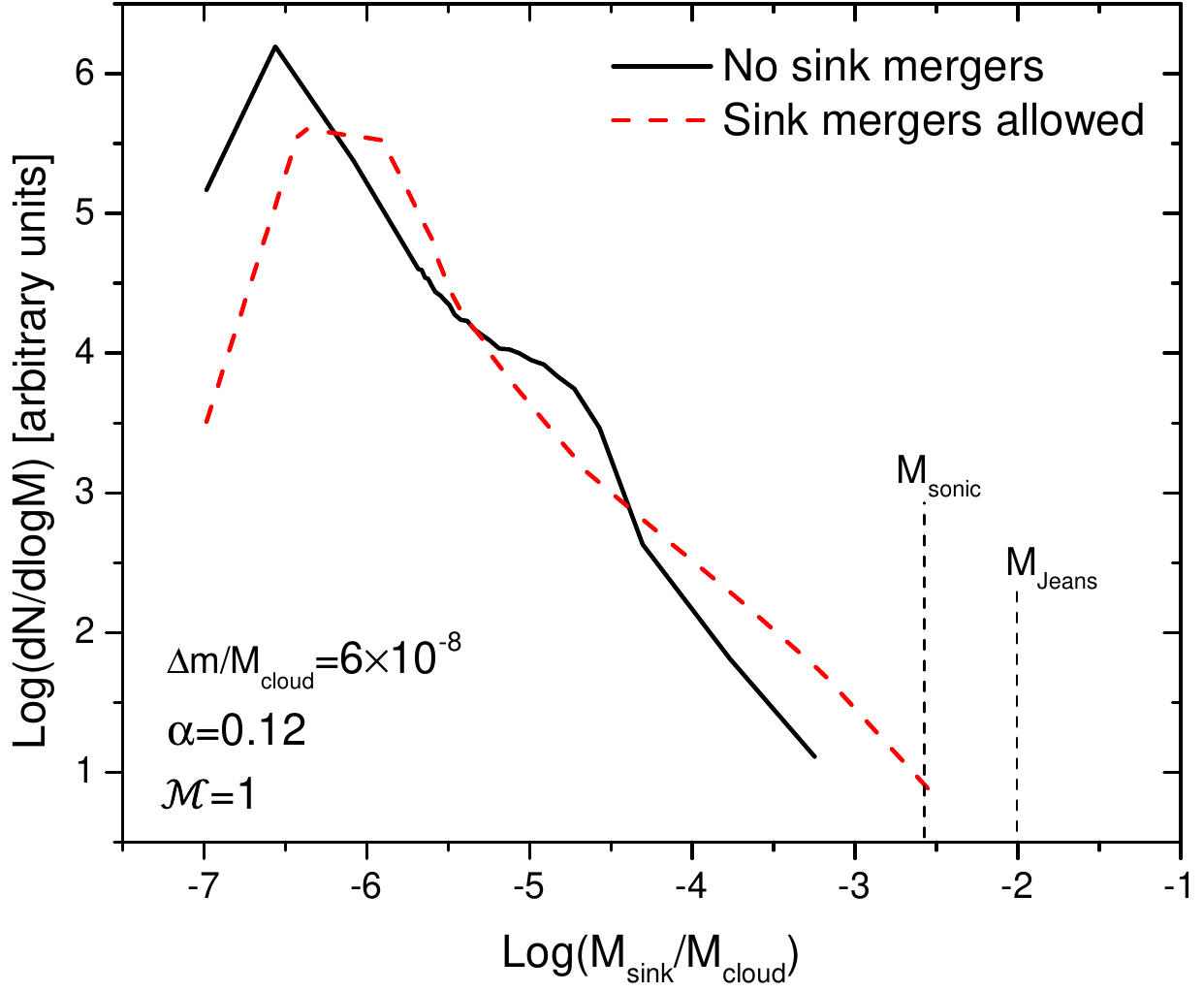}
\caption{The mass distribution of sink particles (IMF), comparing the case where the sink particles are allowed merge (dashed, red) and the case they are not (black, solid). As expected, the overall distribution shifts to larger masses, but the initial conditions (e.g. $M_{\rm Jeans}$ Jeans and $M_{\rm sonic}$ sonic masses) play no role.}
\label{fig:sinkmerger}
\end {center}
\end{figure}

In our simulations sink particles are allowed to merge in order to avoid the spawning of spurious sinks, which can significantly affect their mass distribution. Two sink particles are allowed to merge if the following criteria are met (based on \citealt{Federrath_2010_sink_particle}):
\begin{enumerate}
\item Both are in the same interacting hydrodynamic element.
\item They are gravitationally bound.
\item Their epicentric radius is smaller than 3 times the gravitational force softening and $10^{-4}R_{\rm cloud}$.
\end{enumerate}
To test whether this prescription has any effect on our results we run several simulations where we forbid sink particle mergers. Fig. \ref{fig:sinkmerger} shows that allowing sink particles to merge affects their final mass distribution by decreasing the number of sinks at the resolution limit and thus shifting the peak to a slight higher mass. Overall, it has no qualitative effect on our results as the low-mass cut-off is still determined by the mass resolution.

\end{document}